\documentclass{aa}  

\usepackage{graphicx}
\usepackage{txfonts}
\usepackage[breaklinks=true]{hyperref}
\usepackage{natbib} 
\bibpunct{(}{)}{;}{a}{}{,} 

\usepackage{verbatim}
\usepackage{url}
\usepackage{epstopdf}

\begin{document}

   \title{Uncertainties in grid-based estimates of stellar mass and radius}

   \subtitle{SCEPtER: Stellar CharactEristics Pisa
Estimation gRid}

   \author{G. Valle \inst{1,2}, M. Dell'Omodarme \inst{1}, P.G. Prada Moroni
     \inst{1,2}, S. Degl'Innocenti \inst{1,2} 
          }

   \authorrunning{Valle, G. et al.}

   \institute{Dipartimento di Fisica ``Enrico Fermi'',
Universit\`a di Pisa, largo Pontecorvo 3, Pisa I-56127 Italy
\and
  INFN,
 Sezione di Pisa, Largo B. Pontecorvo 3, I-56127, Italy}

   \offprints{G. Valle, valle@df.unipi.it}

   \date{Received 04/07/2013; accepted 31/10/2013}

  \abstract
   {  
The availability of high-quality astero-seismological data provided by
satellite missions stimulated the development of several grid-based estimation
 techniques to determine the stellar masses and radii.
Some aspects of the systematic and statistical errors affecting these
techniques have still not been investigated well.
}
   {  
 We study the impact on mass and radius determination of the uncertainty in
 the input physics, in the
 mixing-length value, in the initial helium abundance, and in the microscopic
 diffusion efficiency adopted in stellar model computations.  
}
{  
We consider stars with mass in the range [0.8 - 1.1] $M_{\sun}$ and evolutionary
stages from the zero-age main sequence to the central hydrogen exhaustion.  To
recover the stellar parameters, a maximum-likelihood technique was employed by
comparing the observations constraints to a precomputed grid of stellar
models. Synthetic grids with perturbed input were adopted to estimate the
systematic errors arising from the current uncertainty in model computations.
}
  {
We found that the statistical error components, owing to the current typical
uncertainty in the observations, are nearly constant in all cases at about
4.5\% and 2.2\% on mass and radius determination, respectively. 
The systematic
bias on mass and radius determination due to a variation of  $\pm 1$ in 
$\Delta Y/\Delta Z$ is $\pm 2.3 \%$ and $\pm 1.1\%$; the one due to a change
of $\pm 0.24$ in the 
value of the mixing-length $\alpha_{\rm ml}$ is $\pm 2.1\%$ and $\pm 1.0\%$;
the one due to a variation of $\pm 5\%$ in the radiative opacity is $\mp
1.0\%$ and $\mp 0.45\%$. 
An important bias source is to
neglect microscopic diffusion, which accounts for errors of about 3.7\%
and 1.5\% on mass and radius. The cumulative effects of the considered
uncertainty 
sources can produce biased estimates of stellar
characteristics.  Comparison of the results of our technique with other
grid techniques shows that the systematic biases induced by the differences in
the estimation grids are generally greater than the statistical errors
involved.
 }
{}


   \maketitle

\section{Introduction}\label{sec:intro}

Accurate and precise determination of the main stellar parameters is
fundamental in many astrophysical areas. The continuously growing number of
detected extrasolar planets over the past decade has strengthened this
statement even more, since the inferred properties of a planet depend on the mass and
radius estimates of the host star.

The growth of observational asteroseismology through satellite missions, 
such as CoRoT \citep[see e.g.][]{Appourchaux2008,Michel2008,Baglin2009} 
and {\it Kepler} \citep[see e.g.][]{Borucki2010, Gilliland2010},
has opened a new way to estimate stellar properties of solar-type stars, such
as mass, radius, and age. Moreover, these
properties can be determined more precisely by combining these
data with traditional non-seismic observables, such as effective temperature
$T_{\rm eff}$, metallicity [Fe/H], and luminosity $L$.

Different techniques of analysis have been developed to exploit the
growing availability of data. Some of them attempt a direct fit to the
individual 
oscillation frequencies \citep{Metcalfe2009}, while others determine
the stellar parameters by fitting the data to precomputed grid of stellar
models \citep[see
  e.g.][]{Stello2009,Basu2010,Quirion2010,Gai2011,Huber2013}. These 
methods have been recently adopted in stellar population studies 
\citep[see e.g.][]{Chaplin2011,Verner2011,Mathur2012,Miglio2012} because
they allow a fast and automated way to obtain the stellar
characteristics from data.

Although a large amount of work has been done to determine the systematic and
statistical errors inherent in grid-based estimates \citep[see
  e.g.][]{Gai2011, Basu2012}, some aspects have still not been investigated thoroughly. In
particular, the impact of the current uncertainty in the main input
physics required to compute the stellar models at the base of any grid
technique has not yet been quantified fully. In \citet{incertezze1, incertezze2} we showed
that the effect on stellar evolutionary tracks and isochrones of the
uncertainties in the microphysics input (i.e. radiative opacity, nuclear
reaction cross sections, etc.) adopted in stellar codes is not
negligible. Thus, to evaluate the reliability of the values obtained using any
grid technique, it is necessary to understand how the uncertainties in
microphysics propagate into the final results. Besides the input physics, the
stellar models depend on the
 values of some still uncertain parameters, such as the initial
chemical abundances (i.e. the initial helium abundance at a given metallicity)
and the efficiency of the convective transport, i.e. the mixing-length
parameter and the core overshooting
\citep{Chaboyer1995,Barmina2002,Claret2007}.
 
In this paper we address the problem of quantifying the effects of those
uncertainties affecting stellar models on the star properties derived from a
grid technique. We restrict our analysis to central hydrogen-burning stars with
mass in the range [0.8 - 1.1] $M_{\sun}$. We focus on the following
uncertainty sources: the radiative opacity, the microscopic diffusion
velocities, the $^{14}$N$(p,\gamma)^{15}$O reaction rate, the mixing-length
parameter value, and the initial helium abundance-metallicity
relationship. The uncertainty in the convective core overshooting extension is
neglected, since the stars in the chosen mass range burn hydrogen in a
radiative core.

The structure of the paper is the following. In Sect.~\ref{sec:method},
\ref{sec:test} we 
discuss the method and the grids used in the estimate process; 
in Sect.~\ref{sec:griglia} we discuss the effects on the estimates of the
grid morphology; the main
results are presented in Sects.~\ref{sec:results} - \ref{sec:extension}; in
Sects.~\ref{sec:comparison} and \ref{sec:observations}
we present a comparison of the estimates obtained with our grids with some
observations and with those obtained by other techniques. Some
concluding remarks can be found in Sect.~\ref{sec:conclusions}.
 
\section{Grid-based recovery technique}\label{sec:method}

We developed a code -- SCEPtER (Stellar CharactEristics Pisa
Estimation gRid) --  that allows, through a maximum likelihood technique, 
to estimate the stellar mass and radius given a set of observable quantities 
relying on a grid of precomputed stellar models. The code is quite flexible 
since different observables can be used, depending on their availability, 
as well as different grids of models. The code and the grids developed for this
work are available in the R packages
SCEPtER\footnote{\url{http://CRAN.R-project.org/package=SCEPtER}} and
SCEPtERextras\footnote{\url{http://CRAN.R-project.org/package=SCEPtERextras}}
on CRAN.

In the current paper we focus on the case where four quantities are
available: the stellar effective temperature $T_{\rm eff}$, the
metallicity [Fe/H], the large frequency spacing $\Delta \nu$, and the
frequency of maximum oscillation power $\nu_{\rm max}$.  The large frequency
spacing $\Delta \nu$ is the spacings between consecutive overtones having
the same spherical angular harmonic, and they are related to the acoustic
radii of 
the stars \citep[see e.g.][and references therein]{Dalsgaard2012}. The
frequency of maximum oscillation power $\nu_{\rm max}$ is related to the
acoustic cut-off frequency of a star \citep[see e.g.][]{Chaplin2008}. The
selection of these seismic parameters is motivated by their being
recovered from most of target stars.

More in detail, we adopted the same scheme
 described in \citet{Basu2012}. We let $\cal S$ be a star for which the following vector of observed quantities
is available: $q^{\cal S} \equiv \{T_{\rm eff, \cal S}, {\rm [Fe/H]}_{\cal S},
\Delta \nu_{\cal S}, \nu_{\rm max, \cal S}\}$. Then we let $\sigma = \{\sigma(T_{\rm
  eff, \cal S}), \sigma({\rm [Fe/H]}_{\cal S}), \sigma(\Delta \nu_{\cal S}),
\sigma(\nu_{\rm max, \cal S})\}$ be the nominal uncertainty in the observed
quantities. For each point $j$ on the estimation grid of stellar models, 
we define $q^{j} \equiv \{T_{{\rm eff}, j}, {\rm [Fe/H]}_{j}, \Delta \nu_{j}, \nu_{{\rm max}, j}\}$.
Let $ {\cal L}_j $ be the likelihood function defined as
\begin{equation}
{\cal L}_j = \left( \prod_{i=1}^4 \frac{1}{\sqrt{2 \pi} \sigma_i} \right)
\times \exp \left( -\frac{\chi^2}{2} \right)
\label{eq:lik}
\end{equation}
where
\begin{equation}
\chi^2 = \sum_{i=1}^4 \left( \frac{q_i^{\cal S} - q_i^j}{\sigma_i} \right)^2.
\end{equation}
The technique allows an easy expansion whenever other observational constraints
are available, since they can be inserted into the same scheme.

The likelihood function is evaluated for each grid point within $3 \sigma$ of
all the variables from $\cal S$, allowing for a faster computation without
introducing any perturbation on the estimated quantities.  We then let
\begin{equation} 
{\cal L}_{\rm max} = \max_j {\cal L}_j
\end{equation}
be the maximum value of the likelihood function over the estimation grid of
stellar models. The estimated values of radius and mass are obtained
by averaging the radius and the mass of all the models with likelihood greater
than $0.95 \times {\cal L}_{\rm max}$.

Whenever an informative prior on the stellar characteristics to be estimated
is available, the technique can take the information into account in the
likelihood computation. The prior can be inserted as a multiplicative factor
in Eq.~(\ref{eq:lik}), and it can be considered as a weight attached to the grid
points.

The technique can also be employed to construct a confidence interval for mass
and radius estimations.  To this purpose a synthetic sample of $n$ stars is
generated, following a multivariate normal distribution with vector of mean
$q^{\cal S}$ and covariance matrix $\Sigma = {\rm diag}(\sigma)$. A value of
$n = 10000$ is usually adopted since it provides a fair balance between
computation time and the accuracy of the results.  The median of the radius and
mass of the $n$ objects is taken as the best estimate of the true values; the
16th and 84th quantiles of the $n$ values are adopted as a $1 \sigma$
confidence interval.

\subsection{Standard estimation grid}

The standard estimation grid of stellar models is obtained using our well
tested evolutionary code -- FRANEC \citep{scilla2008} -- in the same
configuration as was adopted to compute the Pisa Stellar
Evolution Data Base\footnote{\url{http://astro.df.unipi.it/stellar-models/}} 
for low-mass stars \citep{database2012, stellar}.

The grid consists of 78320 points (110 points for 712 evolutionary tracks),
corresponding to evolutionary stages from the ZAMS to central hydrogen
exhaustion. Models are computed for masses in the range [0.80 - 1.10]
$M_{\sun}$ with a step of 0.01 $M_{\sun}$ and [Fe/H] in the range [$-0.55$ -
  0.55] with a step of 0.05 dex. The solar scaled heavy-element mixture by
\citet{AGSS09} is adopted.  The initial helium abundance is obtained using the
linear relation:
\begin{equation}
Y = Y_p+\frac{\Delta Y}{\Delta Z} Z
\label{eq:YZ}
\end{equation}
with cosmological $^4$He abundance value $Y_p = 0.2485$ from WMAP
\citep{cyburt04,steigman06,peimbert07a,peimbert07b}, and assuming $\Delta
Y/\Delta Z = 2$ \citep{pagel98,jimenez03,gennaro10}. The models are computed
assuming the solar-scaled mixing-length parameter $\alpha_{\rm
  ml} = 1.74$. Present calculations use the most recent version of the OPAL
equation of state, EOS 2006 \citep{rogers1996, rogers02}. 
 Radiative opacity coefficients are taken from the OPAL group
\citep{iglesias96} in the version released in 2006 for temperatures
higher than $10^{4.5}$ K, and from \citet{ferg05} for lower
temperatures. 
Nuclear reaction rates are taken from the NACRE compilation \citep{nacre}
except for $^{14}$N$(p,\gamma)^{15}$O, for which we adopt a more recent
estimate by \citet{14n}. Present models include atmospheric models
by \citet{brott05} as outer boundary conditions. Atomic diffusion is included, taking the
effects of gravitational settling and thermal diffusion into account with coefficients
given by \citet{thoul94}. Further details on the input adopted in the
computations are available in \citet{cefeidi}.

The average large frequency spacing $\Delta \nu$ and the frequency of maximum
oscillation power $\nu_{\rm max}$ are obtained using a simple scaling from
the solar values:
\begin{eqnarray}\label{eq:dni}
\frac{\Delta \nu}{\Delta \nu_{\sun}} & = &
\sqrt{\frac{M/M_{\sun}}{(R/R_{\sun})^3}} \quad ,\\  \frac{\nu_{\rm
    max}}{\nu_{\rm max, \sun}} & = & \frac{{M/M_{\sun}}}{ (R/R_{\sun})^2
  \sqrt{ T_{\rm eff}/T_{\rm eff, \sun}} }. \label{eq:nimax}
\end{eqnarray}
In the case of $\Delta \nu$, a more accurate approach would require the fit of
the pulsational frequencies of each model, but the extraction of frequencies
is generally only possible for high signal-to-noise detections.  Moreover, it
has been shown \citep{Stello2009, Basu2010} that the simple scaled values are
adequate for the grid-estimation process.  For a review of both theoretical
and empirical tests of Eqs. \ref{eq:dni} and \ref{eq:nimax} we refer to
\citet{Belkacem2012, Miglio2013}, and to Sect.~2.1 of \citet{Huber2013}.

\section{Grid technique internal accuracy}\label{sec:test}

To test the consistency of the recovery procedure, we built a synthetic dataset
by sampling $N = 10000$ artificial stars from the same standard estimation
grid of stellar models used in the recovery procedure, adding to each of them
a Gaussian noise in all the observed quantities. We assumed the same standard
deviations as used by \citet{Gai2011}: i.e. 2.5\% in $\Delta \nu$, 5\% in
$\nu_{\rm max}$, 100 K in $T_{\rm eff}$, and 0.1 dex in [Fe/H].

The recovery of mass and radius for these artificial stars by adopting the same
standard grid as was used to generate them will reveal possible distortions
introduced by the technique itself and will serve as reference for all the other
results. This numerical experiment will prove the performances of our
grid technique in the ideal case where the adopted stellar models are in
perfect agreement with real stars.
   
\begin{figure}
\centering
\includegraphics[height=8cm,angle=-90]{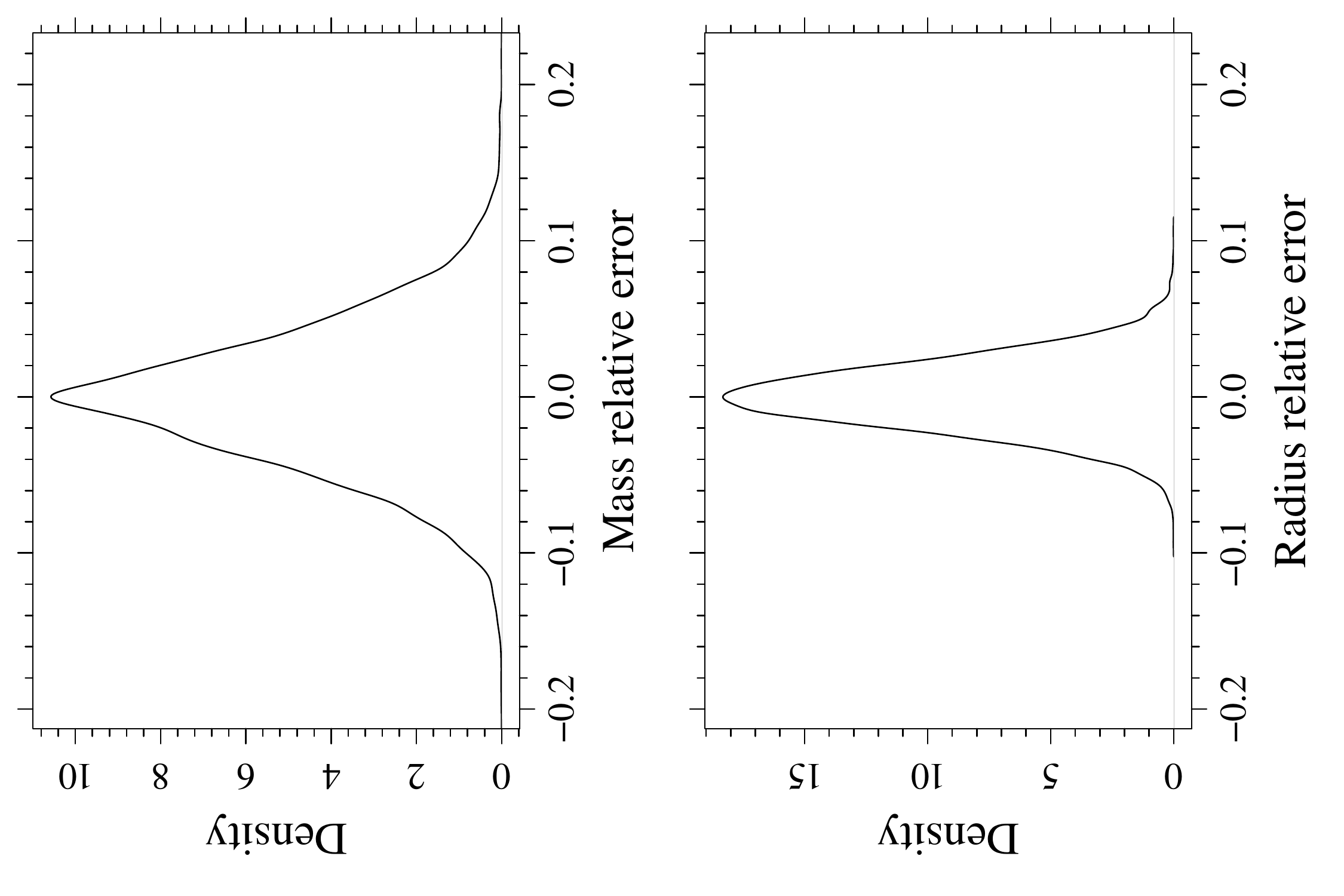}
\caption{(Top) Kernel density estimate of the relative error for mass
  reconstruction (bandwidth 0.0060). (Bottom) As the upper panel but for the
  radius (bandwidth 0.0031). Stars are sampled and reconstructed on the
  standard grid. A positive value of the
  relative error indicates an estimated value greater than the true value.}
\label{fig:std-grid}
\end{figure}

In Fig.~\ref{fig:std-grid} we present the kernel density estimates of the
relative error for the mass and radius reconstruction \citep{Scott1992,
  venables2002modern}. Some details on the adopted technique  are given in
Appendix~\ref{sec:smooth}.  In all the cases a positive value of mass (radius)
relative error indicates an estimated value greater than the true value. The
relevant quantities of the distributions are summarized in rows 1 (labelled
``standard'') of Table \ref{tab:results}. In the table we report, for mass and
radius relative errors, the median of the distribution; the width of its 95\%
confidence interval; the standard deviation of the data; the width of its 95\%
confidence interval; the 16th and 84th quantiles of the data. For ease of
identification, each case is identified by a number and a label, which are
identical for mass and radius.

Since the median of the relative error of mass and radius estimates are
consistent with zero, the technique is unbiased for the standard case
(i.e. the technique is accurate). The standard deviation values for the 
mass and radius estimates are of the order of, respectively, 4.5\% and 
2.2\%, so the method provides good recovery precision; however, the error on the single estimate can be as much as 20\% and 
10\% for mass and radius, respectively.

Table~\ref{tab:results-rederror} gives the standard deviation of the relative
errors in the mass and radius estimates for different assumptions on the
adopted errors in the observed quantities. This allow to explore the
sensitivity of the reconstruction procedure to the precision degree of the
observational data. As a first test, lines 1 in
Table~\ref{tab:results-rederror} show the effect of doubling the precision of
effective temperature measurements (i.e. 50 K as $T_{\rm eff}$ error) and
keeping the uncertainties in the other
quantities fixed to the standard value. As a second test, lines 2 in Table~\ref{tab:results-rederror} show
the impact of halving the error in metallicity measurements (i.e. 0.05 dex as
[Fe/H] error), again adopting the standard value for the uncertainty in the
other quantities. The third test assumes 1\% in $\Delta \nu$ and 2.5\% in
$\nu_{\rm max}$ (lines 3 in Table~\ref{tab:results-rederror}). The last test
(lines 4 in Table~\ref{tab:results-rederror}) assumes the set of all reduced
errors quoted above.  

Concerning mass reconstruction, increasing the $T_{\rm eff}$ precision of a factor
of two has a major effect on the reconstructed mass, as it leads to a
reduction  
of the standard deviation on relative errors from 4.5\% to 3.5\%.  The second
most   
important effect is due to improving the metallicity measurements, whereas
having more  
precise $\Delta \nu$ and $\nu_{\rm max}$ only leads to a minor improvement. 
This behaviour can be understood by considering the mass gradient with 
respect to $T_{\rm eff}$  and seismic parameters in the neighbourhood of
a reference point. It turns out that the dependence of mass on $T_{\rm
  eff}$ is much steeper than the one on seismic parameters. The stellar tracks for 
  the evolutionary stages considered in this paper evolve at slightly changing
    $T_{\rm  eff}$, but rapidly decreasing seismic parameters (see e.g. the
  right panel 
in Fig.~\ref{fig:onlyDnu-grid-analisi}).

Regarding radius determination, increasing the precision in the two
asteroseismic observables 
 has the strongest impact, reducing the
standard deviation on relative errors from 2.2\% to 1.6\%. More precise
effective temperature  
and metallicity measurements only lead to a small improvement in radius
precision.  
All these results agree with those of \citet{Basu2012}.
The set of all reduced errors roughly halves the standard deviation of the
mass and radius 
estimates.

\begin{figure*}
\centering
\includegraphics[width=11cm,angle=-90]{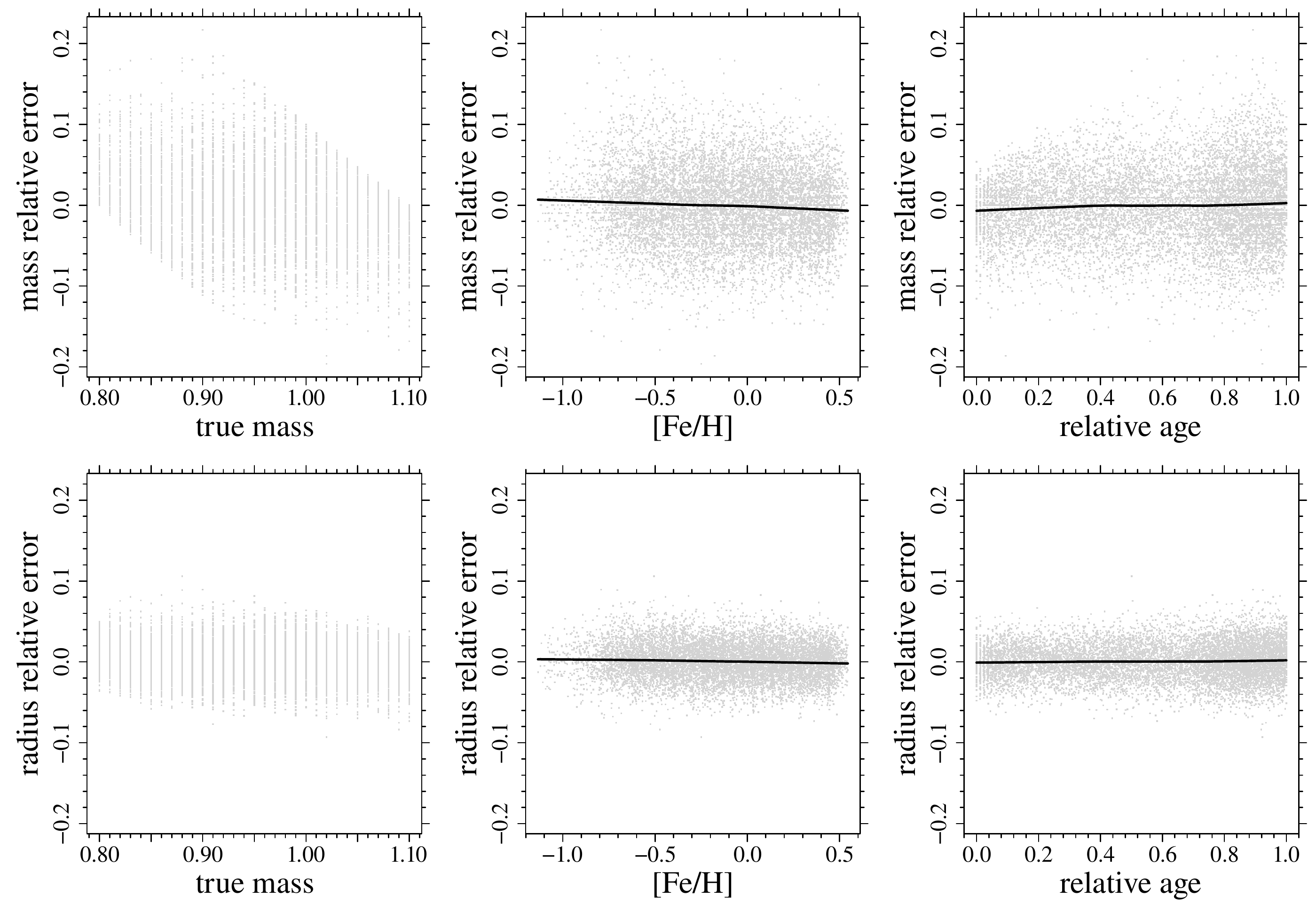}
\caption{(Upper row) Dependence of mass estimate relative errors on
  the true mass of the star, on the metallicity value, [Fe/H], and on the
  relative age of the star (see text). The [Fe/H] is the surface value
  currently present for the star, not the original one. (Lower row) The same
  as upper row but for
  radius reconstruction. Data are sampled and reconstructed on the standard
  grid of stellar models. The black lines represent a LOWESS smoother of the
  data.  
  A positive value for the  
  relative error indicates an estimated value greater than the true value.}
\label{fig:std-grid-analisi}
\end{figure*}

Figure~\ref{fig:std-grid-analisi} shows the dependence of relative errors on
the recovered mass and radius on the true stellar mass, [Fe/H] value, and
relative age, defined as the ratio between the age of the star and the age of
the same star at central hydrogen exhaustion. (The age is conventionally set to
0 at the ZAMS position.) The ratio [Fe/H] is the current surface value 
for the star, which can be significantly different from the
initial one owing to the microscopic diffusion processes. The figure shows  a
LOWESS (LOcally Weighted Scatterplot Smoothing) smoother of the 
data \citep{Cleveland1981}. More details on the technique are given in
Appendix~\ref{sec:smooth}.  

The most evident feature is the "edge effect"
that characterizes the trend of the mass relative errors versus the true mass
of the stars. The trend is a distinctive feature of a maximum likelihood grid
technique, 
although its relevance is not always recognized. This is because
the mass can never be estimated at values outside the
grid, so that the estimate of the mass of a star with true mass of 0.80
$M_{\sun}$, the lowest value available in the grid, can never result in values
below 0.80 $M_{\sun}$. The opposite occurs  for a star of true mass 1.10
$M_{\sun}$, our highest value; i.e. the estimated mass
 can never result in a value greater than 1.10 $M_{\sun}$. As a
consequence the apparent precision of mass estimate on the grid is higher
toward its 
edges, but these estimates are systematically biased. Very
similar behaviour is 
reported, for instance, in \citet{Gai2011} (their Fig.~9) although its causes
are not discussed. 

As shown in the left-hand panel of the bottom row in
Fig.~\ref{fig:std-grid-analisi}, the same effect, although less evident,
 is present in the decreasing trend of  the radius relative errors versus the
 true mass of the stars.  
In this case the trend is induced by the fact that the radius of a star
increases with the star mass, so that for the lowest mass (0.80 $M_{\sun}$) 
the probability of having an estimate of radius below the edge of the grid is
null, while it is possible to have radius estimate above the grid values
spanned by 0.80  $M_{\sun}$ models, because of the increasing trend mentioned above.
The reverse behaviour occurs for 1.10 $M_{\sun}$ models. A more detailed 
discussion of the edge effects can be found in the next section. 

Interestingly, no clear trend is shown for different values of [Fe/H].
The region at [Fe/H] < $-0.8$ dex is poorly populated since
 the microscopic diffusion can reduce the surface [Fe/H] at these
values only for a few tracks that start at low metallicity. 
For the dependence on relative age, a mild increase in the
standard deviation is found for the mass estimate: from a value of
0.037 for relative ages in the range [0.00 - 0.25] to a value of 0.049 in the
range [0.75 - 1.00]. This is another example of the edge effect, a pervasive
presence in several aspects of the grid-based estimation.
More details on this topic are
given in Sect.~\ref{sec:griglia}.

As a last comment, 
as reported in the literature \citep[see e.g.][]{Jorgensen2005}, 
neglecting the evolutionary time step in grid-based calculations
can lead to significant biases in the estimated stellar parameters. However,
the mass range adopted in our computations -- which stops at 1.10
$M_{\sun}$ -- including metallicity and seismic constraints and 
excluding the PMS and post central hydrogen exhaustion 
phases, reduce the occurrence of intersecting tracks. To quantify the
bias,
we performed a weighted grid estimate of mass, using the evolutionary
age as weight. Only a small median bias of -0.003 $M_{\sun}$, which is constant for all ages,
is found. No relevant
differences appear for the relative error in mass estimates.

\section{Trends induced by the grid morphology}
\label{sec:griglia}

In Sect.~\ref{sec:test} a mild trend in the standard
deviation of the relative errors on mass estimates with the relative age of the
star is reported . Although the effect is very small for most practical purposes, it
deserves analysis since it is typical of grid-based estimation techniques.
Such an effect is related to the morphology of the projection of
the grid in the ($T_{\rm eff}$, $\Delta \nu$) and in the ($T_{\rm eff}$,
$\nu_{\rm max}$) planes. To keep the discussion simple and to increase the
relevance of the effect under investigation, we focus on the case where
only $T_{\rm eff}$ and $\Delta \nu$ (or equivalently $\nu_{\rm max}$) are
available.

\begin{figure*}
\centering
\includegraphics[height=18cm,angle=-90]{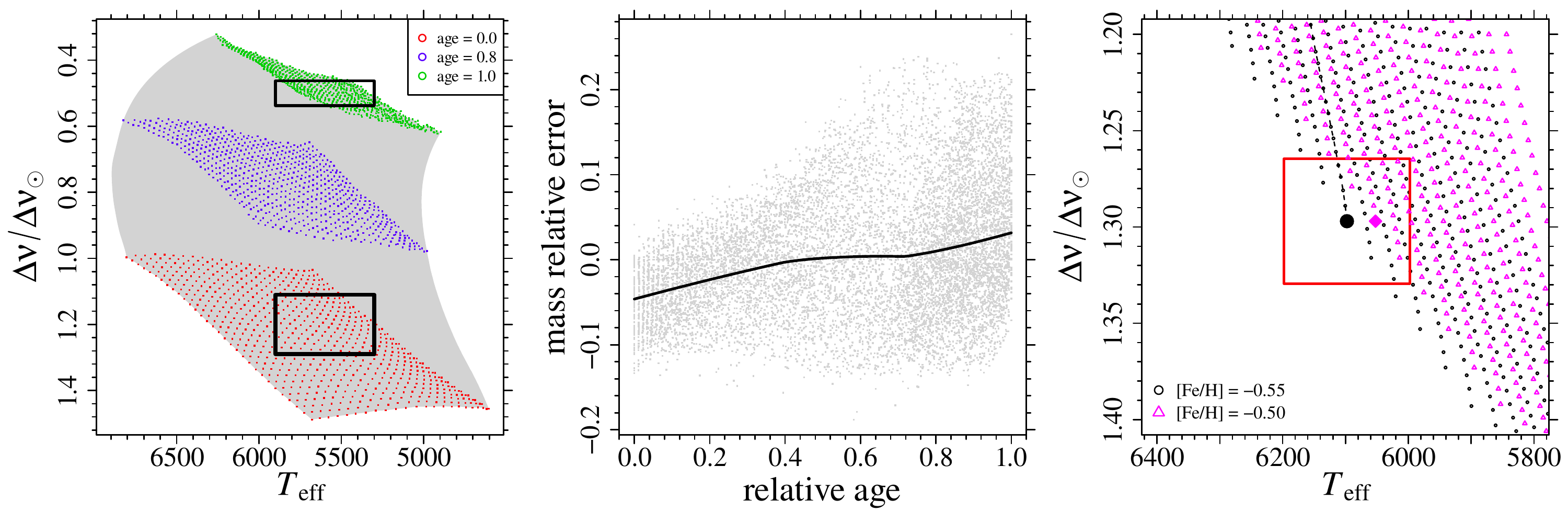}
\caption{(Left): region of the  ($T_{\rm eff}$, $\Delta \nu$)
  plane spanned by the reconstruction grid; the points mark the position of
  ZAMS (red), of the relative evolutionary age 0.8 (blue), and of the central
  hydrogen 
  depletion (green). The black boxes show the typical extent of the $3 \sigma$
  regions in the initial and final stages of the stellar evolution.   
  (Middle): relative
  error on mass estimates with respect to the relative age of the stars. Only
  $T_{\rm eff}$ and $\Delta \nu$ are employed for the estimate. (Right): detail
  of the region around the $M$ = 0.90 $M_{\sun}$ (black dot) for initial 
  [Fe/H] = $-0.55$. The 
  red box shows the extent of $1 \sigma$ region around the aforementioned
  model. The magenta diamond corresponds to the same model for initial [Fe/H] =
  $-0.50$. The dashed line marks the evolutionary path of the model.
}
\label{fig:onlyDnu-grid-analisi}
\end{figure*}

In the left-hand panel of Fig.~\ref{fig:onlyDnu-grid-analisi}, we plot the
projection of the estimation grid in the ($T_{\rm eff}$, $\Delta \nu$)
plane. The evolutionary path of a star starts at the bottom edge of the grid and
develops towards lower values of $\Delta \nu$. In the figure we display the
position of the ZAMS points, of the points that mark the 80\% of the
evolution, and of the final points corresponding to the central hydrogen
depletion.  

The figure shows that during the evolution of a MS star, a slight variation in$T_{\rm eff}$ 
corresponds to a large decrease in $\Delta \nu$. It also appears that the
points corresponding to  
the models with relative ages 0.0 and 0.8 are more scattered in $\Delta
\nu$ than those with relative age 1.0. Moreover, the separation in
  $\Delta \nu$  
of regions at given relative ages increases as a star evolves from the
ZAMS towards  
the central-hydrogen exhaustion. In fact, as shown in the left-hand panel of
Fig.~\ref{fig:onlyDnu-grid-analisi}, the  
$\Delta \nu$ decrease from the ZAMS to the models with relative age 0.8 is
nearly the same of the decrease from the model
with relative age 0.8 to  
the one corresponding to central-hydrogen exhaustion. Furthermore, the
dimension  
of the $3 \sigma$ boxes around a point shrinks in $\Delta \nu$ at
later ages, since the error on seismic quantities is assumed to be a fixed
percent of the seismic values. This effect is shown in the figure.

In the middle panel of Fig.~\ref{fig:onlyDnu-grid-analisi} we display the
relative error on mass estimates versus the relative age of the stars.  The
increase in the variance with relative age is much more evident than in the
case discussed in Sect.~\ref{sec:test}, where the availability of the
metallicity on 
the estimation procedure mitigates the effect.
The trend is due to the different probability of selecting an object of mass
higher or lower than the one of the sampled object in the maximum likelihood
analysis. This is shown in the right-hand panel of
Fig.~\ref{fig:onlyDnu-grid-analisi}.  
 It is clear that
most of the models encompassed by the $1 \sigma$ box correspond to mass lower (at lower
$T_{\rm eff}$) than the reference case one. Moreover, tracks
  corresponding to lower masses enter in the box at later stages of evolution
  than tracks corresponding to higher masses. 
The conclusion holds even if
tracks at different metallicity are included.
As a result the mass is generally underestimated for low values of relative
age. To help estimate the effects, we note that the relative age of 0.20 is
reached at around the 18th point of the track.

As we move to higher relative
ages, the mass underestimation disappears, and the standard deviation of the mass
relative error estimates increases.
To understand this increase we refer to the case treated in
Sect.~\ref{sec:test} and 
assume that the metallicity of the star is available, along with $T_{\rm eff}$
and seismic parameters. We selected the reference points
corresponding  
to target relative ages 0.0, 0.4, 0.8, and 1.0 in the grid. We then considered all the models in the
$3 \sigma$ boxes of the reference points and computed, for the four
reference ages separately, the quantity $\Delta M$ defined as the difference
between the mass of the reference point and the mass of the models in the
box.

\begin{figure*}
\centering
\includegraphics[height=9cm,angle=-90]{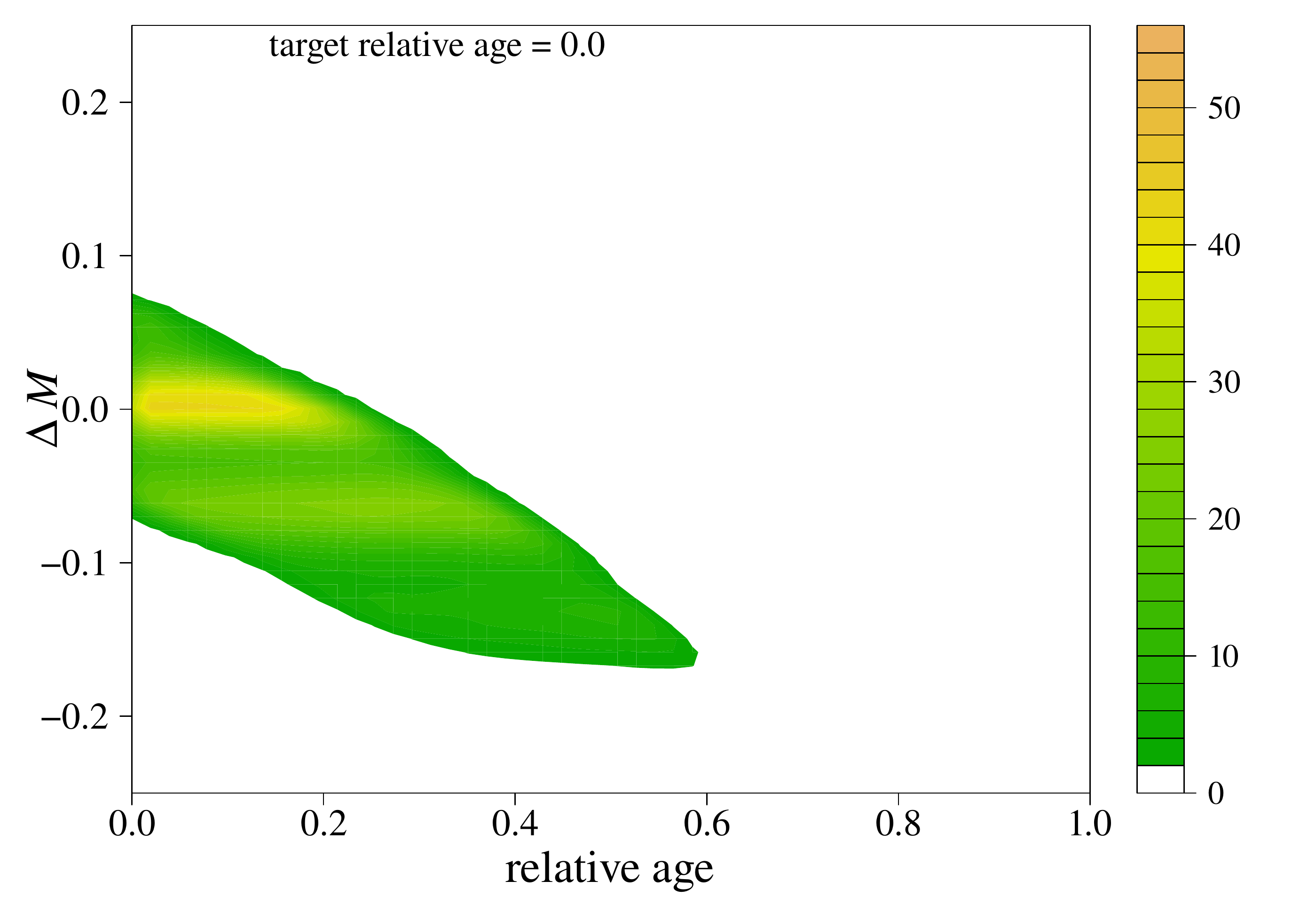}
\includegraphics[height=9cm,angle=-90]{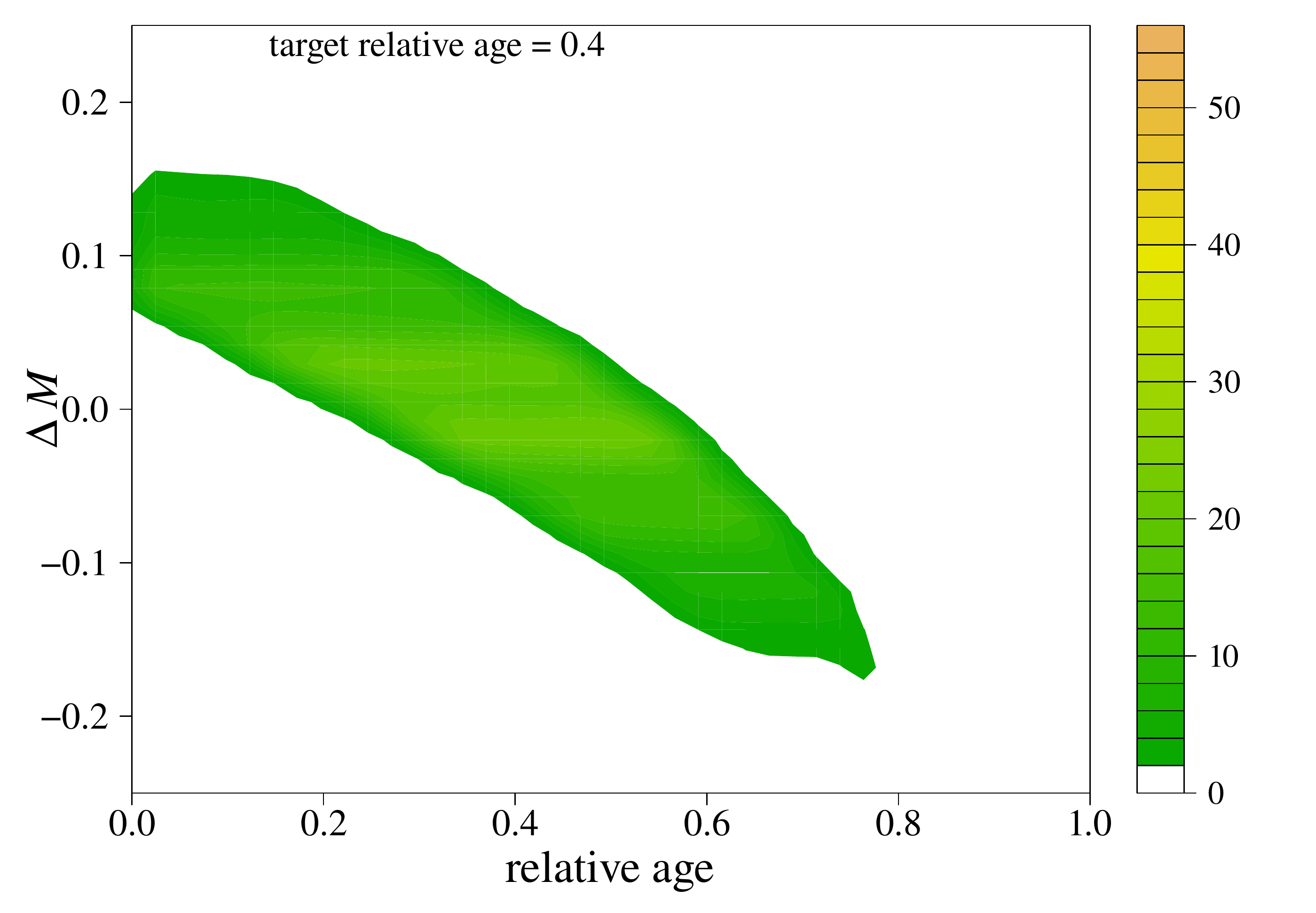}\\
\includegraphics[height=9cm,angle=-90]{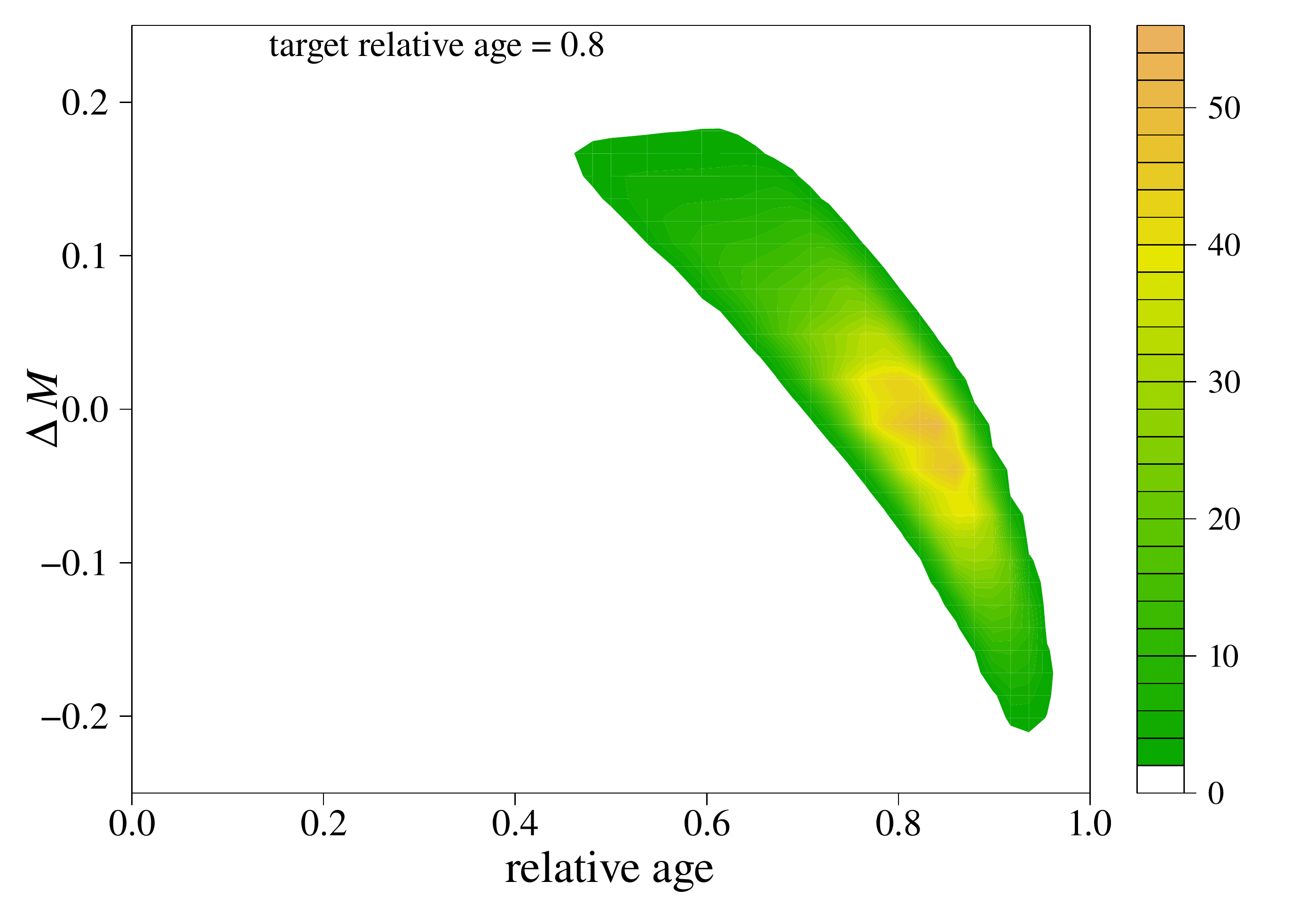}
\includegraphics[height=9cm,angle=-90]{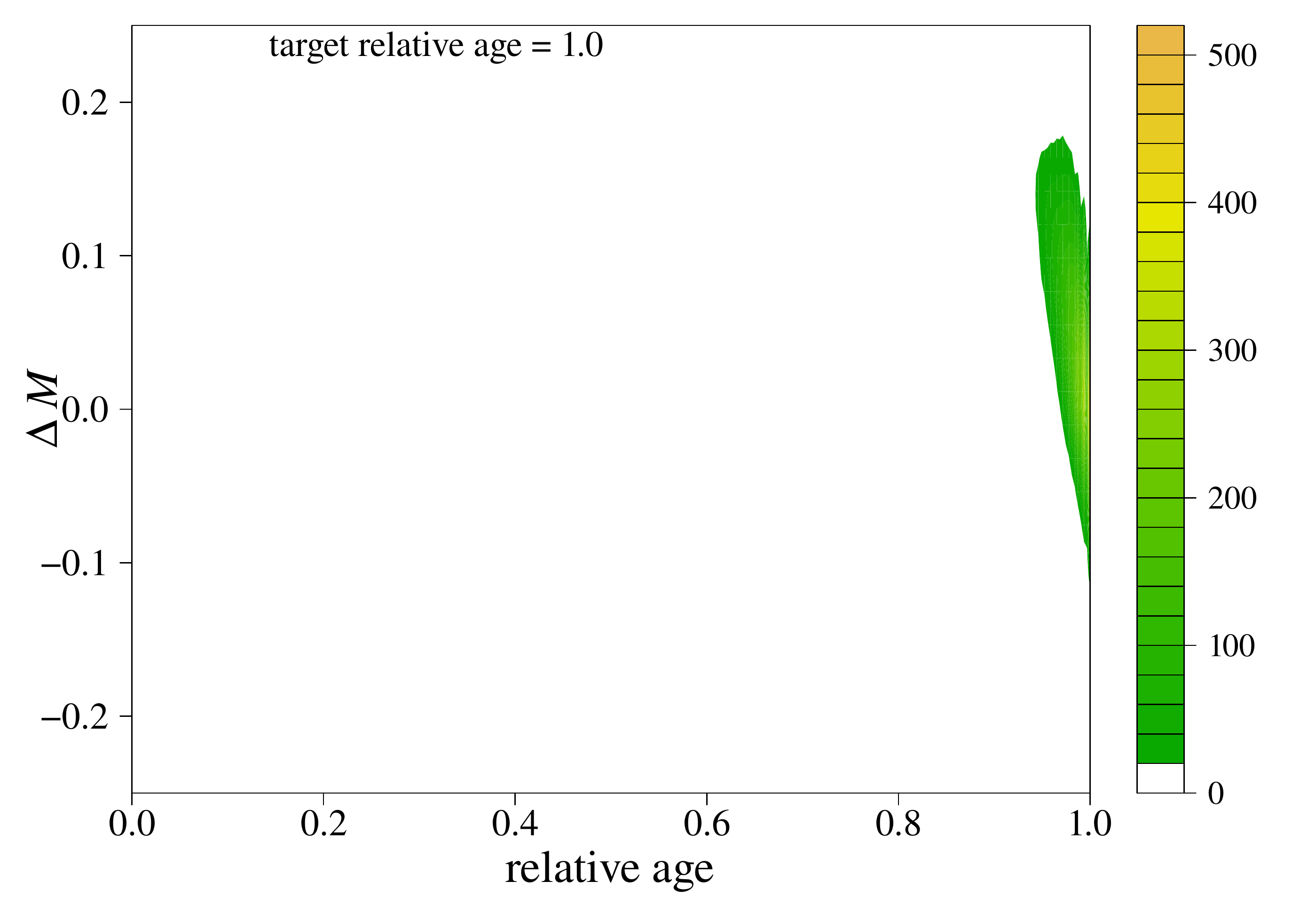}
\caption{Joint density of $\Delta M$ and relative ages selected in the $3
  \sigma$ boxes around the references point for target relative ages 0.0, 0.4,
  0.8, and 1.0. A 
positive value of $\Delta M$ implies that the selected mass is greater than
the reference one. The color legend of the lower right panel is
different from the others because the distribution is much more
peaked.} 
\label{fig:joint-age-M}
\end{figure*}

In Fig.~\ref{fig:joint-age-M} we plot 
the joint density of $\Delta M$ and relative age of all the models at the four
selected target relative ages. 
The density of the extreme cases
(target relative age 0.0 and 1.0) are strongly peaked in mass, in particular
for the case  
at late age. It is also apparent the effect induced by the edge of the grid,
mainly at target relative age 0.0, where the distribution toward higher masses
is truncated at the grid edge.
In the middle and third quarters of the evolution (i.e. relative ages 0.4 and
0.8, respectively),  
the mass densities are flatter, so the variances are higher than in the
previous two cases.  
In other words, at intermediate relative ages, the grid is populated by models
covering  
a mass range  that is wider than in the other two cases.

The trend and the asymmetries of Fig.~\ref{fig:joint-age-M} are the
consequences  of stellar evolution. The changes are smooth in the first part
of the 
evolution; the seismic quantities evolve faster in
the later stages, after about relative age 0.8.  
In fact, as discussed above, in the first three quarters of the evolution,
the reference points in each reference level span a wide
hyper-range, and the levels are close one another (see left panel,
Fig.~\ref{fig:onlyDnu-grid-analisi}). At 
late evolutionary stages, they pack more
closely in each levels and separate strongly from one another. 
As a consequence, at late relative ages, a $3 \sigma$ box
around the reference point contains more homogeneous grid models, 
i.e. models with more similar masses and relative ages than in the 
early evolution.

\section{Stellar-model uncertainty propagation}\label{sec:results}

The accuracy
and precision of the parameters of real stars inferred by means of any
grid-based technique depend on the reliability degree of the adopted stellar
models.  The evolutionary tracks, hence the grid-based results, are
susceptible to variations arising from the adopted input physics
(radiative opacity, nuclear reaction cross-sections, etc.), efficiency of
macroscopic processes (e.g. convection), and initial chemical composition.

For the first issue, we have recently shown that the cumulative
uncertainty affecting the current generation of stellar models from the
combined effect of the main input physics is still not negligible \citep[see
  e.g.][for a detailed discussion]{incertezze1, incertezze2}.
Regarding the second point, it is well known that the rough treatment of
super-adiabatic convective transport is one of the major weakness in stellar
computations. This means that the predicted effective temperature of stars
with a convective envelope strongly depends on the adopted value of the
mixing-length parameter.
Finally, stellar tracks and isochrones clearly depend on the initial chemical
composition adopted in the computations.
  
The variations of the stellar tracks due to all the uncertainty sources
mentioned above will affect the estimates obtained from grid techniques in a
non-trivial way, owing to the concurrent effects arising from a single
ingredient variation.  Direct estimates obtained from perturbed stellar
models are the only way to tackle the problem.  

To this purpose we computed
several sampling grids of non-standard stellar tracks, with mass steps of 0.02
$M_{\sun}$, adopting perturbed input physics, different values of the helium
abundance at a given metallicity, and various mixing-length parameters
$\alpha_{\rm ml}$.  From these non-standard grids we sampled the corresponding
synthetic datasets of $N = 10000$ artificial stars. We applied to these
datasets our recovery procedure based on the standard grid of stellar models 
in order to estimate mass and radius of the artificial objects as in the case of
real 
data. Comparing these reconstructed values with the known true ones allows the effect of the various uncertainty sources discussed in detail to be quantified in
the following sections.

\subsection{Initial helium abundance}
\label{sec:he}

The observable quantities of a star of given age and mass depend on its
chemical composition. While the surface metallicity can be measured, the
helium abundance cannot be determined for the vast majority of stars, since
helium lines are not observable in stars colder than about 20000 K. From the
theoretical point of view, this means that one has to assume an initial value
$Y$ for the helium abundance to adopt in stellar evolution computations. The
common procedure consists in assuming the linear relationship between the
original helium $Y$ and metallicity $Z$ shown in Eq.~(\ref{eq:YZ}). However, the
value of the helium-to-metal enrichment ratio $\Delta Y/\Delta Z$ is still
quite uncertain \citep{pagel98,jimenez03,gennaro10}, since its determination
relies only on indirect methods. Such an uncertainty directly translates into
an uncertainty in the initial helium abundance adopted in stellar computations
for a given initial metallicity and, in turn, into an uncertainty in the
predicted observable quantities of a star.
 
\begin{figure}
\centering
\includegraphics[height=8cm,angle=-90]{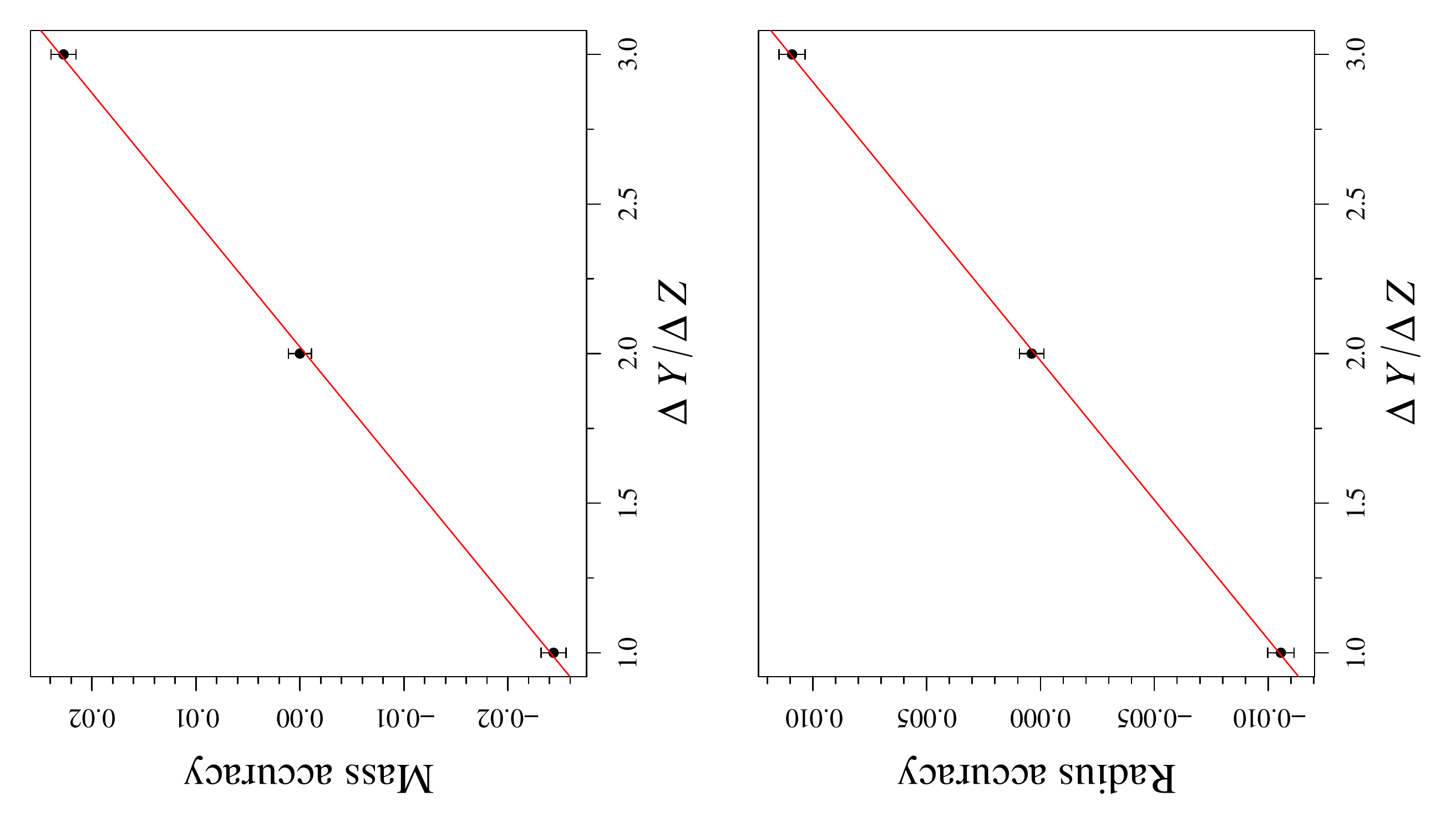}
\caption{Median of mass and radius relative errors; the error bars correspond
  to their 95\% confidence interval. The synthetic data are sampled from grids 
  of stellar models with $\Delta Y/\Delta Z$ = 1, 2, 3 and reconstructed on
  the standard grid 
  with $\Delta Y/\Delta Z$ = 2.}
\label{fig:median-He}
\end{figure}

To quantify the impact of a change in the star's initial helium
content on the estimate of its mass and radius, we computed two additional
grids of stellar models with the same values of the metallicities $Z$ as in
the standard grid but by changing the helium-to-metal enrichment ratio in
Eq.~(\ref{eq:YZ}) within the current uncertainty, namely $\Delta Y/\Delta Z$ =
1 and 3. Then, we built a synthetic dataset of $N = 10000$ artificial stars
by sampling the objects from the non-standard grid with $\Delta Y/\Delta Z =
1$ and another from the non-standard grid with $\Delta Y/\Delta Z = 3$. The
mass and radius of the objects are then estimated using the recovery procedure
based on the standard grid. The results of these tests are presented in rows
2-3 of Table~\ref{tab:results} and in Fig.~\ref{fig:median-He}. The figure
shows the values of the median of the relative error in mass and radius
estimates, where the error bars show their 95\% confidence interval. The lines
represent the weighted least squares fit to the values.
 
It appears that the effect of changing the initial helium abundance is highly
symmetric 
around the standard value, which corresponds to the case discussed in
Sect.~\ref{sec:test}. The bias induced by the considered uncertainty in initial
helium abundance is of the order of $ \pm $2.3\% on mass estimates and $ \pm
$1.1\% on the 
radius.  As expected, an increase (decrease) in the initial helium content of
the stars shifts the mass estimate to higher (lower) values. In fact, 
with respect to the standard grid models, the helium-rich ones have higher
effective temperature and radius, and 
consequently lower seismic parameters. Recalling that as the mass increases,
$T_{\rm eff}$ increases, whereas seismic parameters decrease, it follows that
the mass of the 
sampled object will generally be overestimated.
In this case the grid technique shows a bias, although the statistical errors,
represented by the standard deviation, are still dominant on average (but see
e.g.  
discussion below on metal-rich stars).

\begin{figure*}
\centering
\includegraphics[width=11cm,angle=-90]{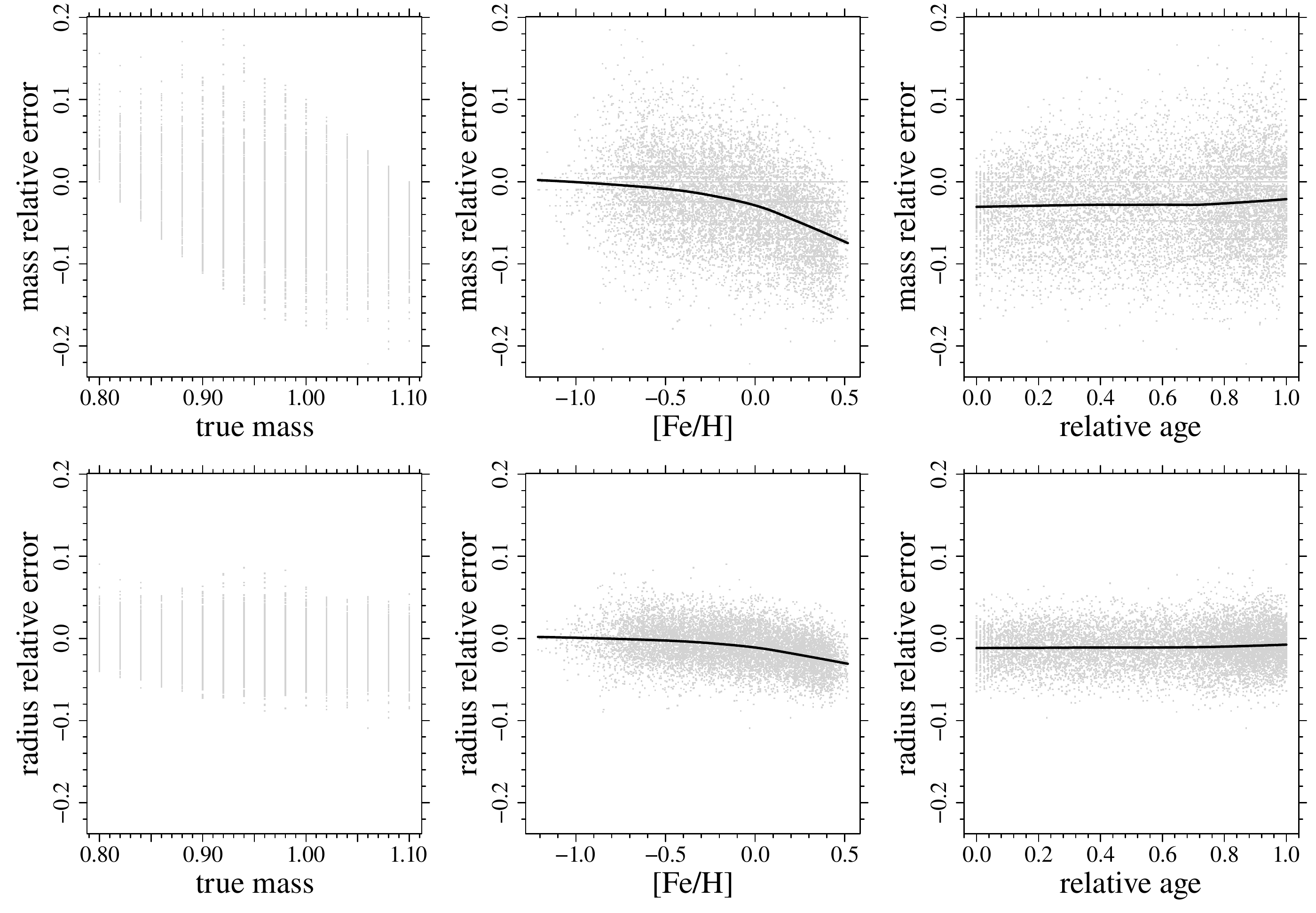}
\caption{As in Fig.~\ref{fig:std-grid-analisi}, but for 
data sampled from a grid with $\Delta Y/\Delta Z = 1$,
  and recovered with the standard grid adopting $\Delta Y/\Delta Z = 2$.}
\label{fig:He-grid-analisi}
\end{figure*}

Figure~\ref{fig:He-grid-analisi} shows the same quantities as
Fig.~\ref{fig:std-grid-analisi} but for the case of synthetic data sampled from 
stellar models adopting $\Delta Y/\Delta Z = 1$. 
Apart from the trend with mass discussed above, there
  is a signature of an edge effect in the trends on relative age, similar
  to the one discussed in Sect.~\ref{sec:griglia}. 
The most interesting feature is, however, a clear metallicity
effect.  This is because $Y$ and $Z$ are linked by the relation
in Eq.~(\ref{eq:YZ}) so that at low $Z$ values the change in $Y$ driven by a
change in $\Delta Y/\Delta Z$ is less than the same change at high
metallicity.

As a result, the uncertainty in helium content mainly affects stars
at high values of [Fe/H]: the median of the distribution of mass relative
errors is $-0.005$ for [Fe/H] $\leq$ $-0.5$, $-0.016$ in the range of [Fe/H] =
[$-0.5$, 0.0], and it grows to a value of $-0.048$ for [Fe/H] $\geq$ 0.0.  The
last error is comparable to the standard deviation of mass relative-error
distribution.

The effect of helium uncertainty should then be carefully
taken into account if the technique is applied to metal-rich objects, since it
can result in biased estimates of mass and radius. The trend discussed above
is reversed whenever the initial helium content is increased with respect to
the standard grid (plots with $\Delta Y/\Delta Z = 3$ not shown).

\subsection{Mixing-length value} 
\label{sec:ml}

The lack of a solid and fully consistent treatment of the convective transport
in superadiabatic regimes prevents modern stellar evolution codes to firmly
predict the effective temperature and radius of stars with an outer convective
envelope, such as those studied in this paper. The common approach consists in
adopting a simplified treatment as the mixing-length theory, where the
efficiency of the convective transport depends on a free parameter
$\alpha_{\rm ml}$ to be calibrated with observations.  Usually, stellar models
adopt the solar calibration that in our standard case provides $\alpha_{\rm
  ml} = 1.74$. Nevertheless, there is no stringent a priori reason that
guarantees that the solar calibrated $\alpha_{\rm ml}$ value is also suitable for stars of different masses and/or in different evolutionary stages.

To establish the influence of varying the super-adiabatic
convective transport efficiency, we computed four additional grids of stellar
models by 
assuming mixing-length parameters that are different from the solar-calibrated
one: $\alpha_{\rm ml}$ = 1.50, 1.62 1.86, and 1.98. Then, we built four
different synthetic datasets of artificial stars by sampling $N = 10000$ stars
for each of the four supplementary non-standard grids with different values of
the mixing-length parameter. Then we applied our recovery procedure, based on
the standard grid of stellar models, on these datasets in order to estimate the
stellar masses and radii. The comparison between recovered and true values
allows the effect of a variation in the mixing-length parameter to be quantified.

\begin{figure}
\centering
\includegraphics[height=8cm,angle=-90]{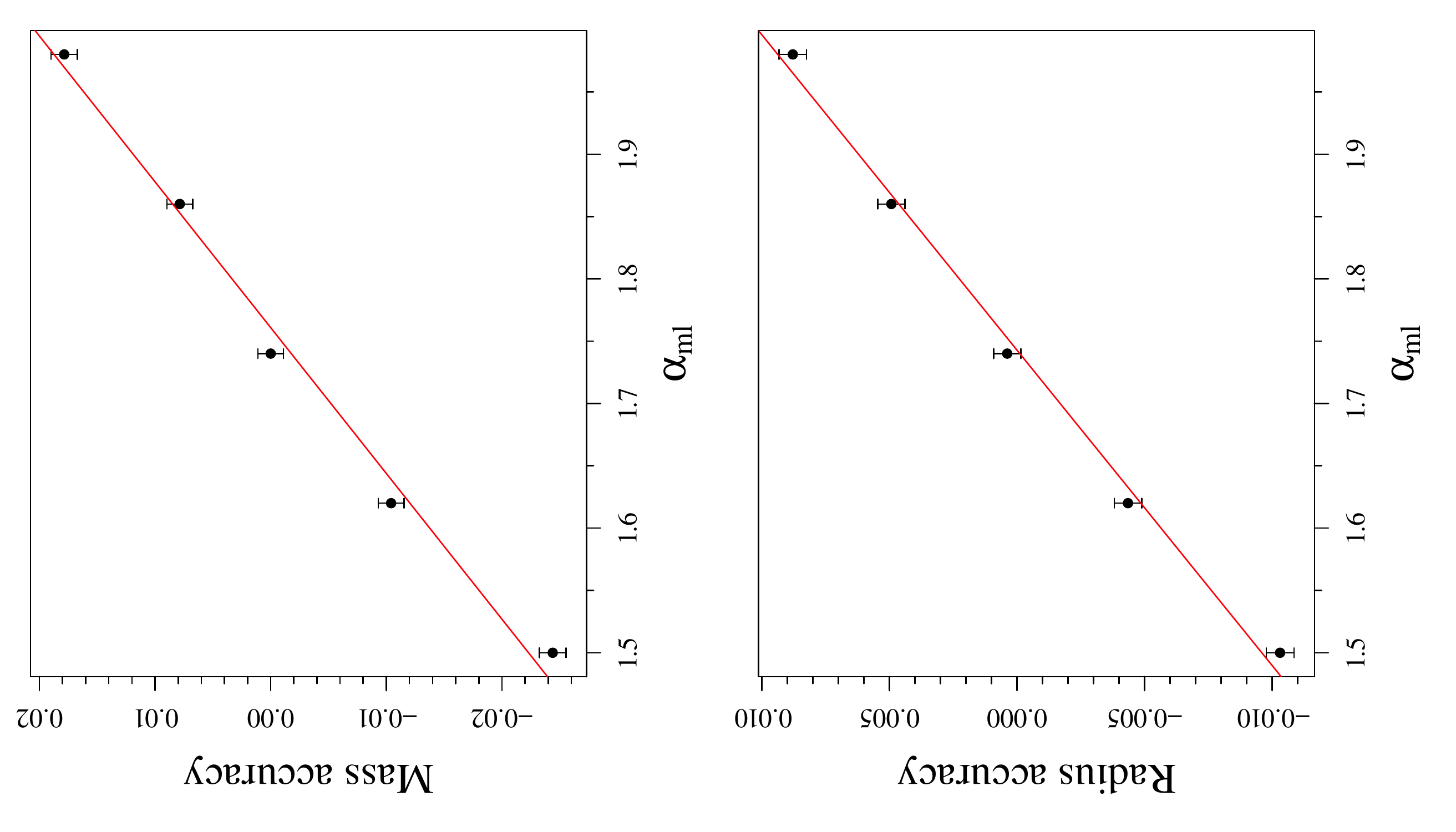}
\caption{As in Fig.~\ref{fig:median-He}, but 
for synthetic data sampled from grids with different $\alpha_{\rm ml}$ values
  and reconstructed with the standard grid for
  $\alpha_{\rm ml} = 1.74$.}
\label{fig:median-ML}
\end{figure}

The results are shown in rows 4-7 of Table~\ref{tab:results} and in
Fig.~\ref{fig:median-ML} for both mass and radius estimates. The extreme
change in $\alpha_{\rm ml}$ (i.e. $\pm  0.24$) induces a bias in mass and 
radius estimates of about $\pm  2.1\%$ and $\pm  1.0\%$, which is similar to
the one caused by the $\Delta Y/\Delta Z$ variation discussed in 
Sect.~\ref{sec:he}. Figure~\ref{fig:median-ML} shows the dependence of the median
of the relative errors distribution on the $\alpha_{\rm ml}$ adopted in the
sampling grids. The trend is nearly linear, although the fit is not as good as
in the other cases discussed in this work.  As for initial helium abundance
variation, the bias is dominated by the statistical errors. 

As expected, the synthetic datasets sampled from the models with higher
(lower) values of $\alpha_{\rm ml}$ are reconstructed with higher (lower)
masses.  In fact a higher value of $\alpha_{\rm ml}$ results in MS synthetic
stars with higher effective temperature, which mimic more massive objects
 with standard $\alpha_{\rm ml}$.

\begin{figure*}
\centering
\includegraphics[width=11cm,angle=-90]{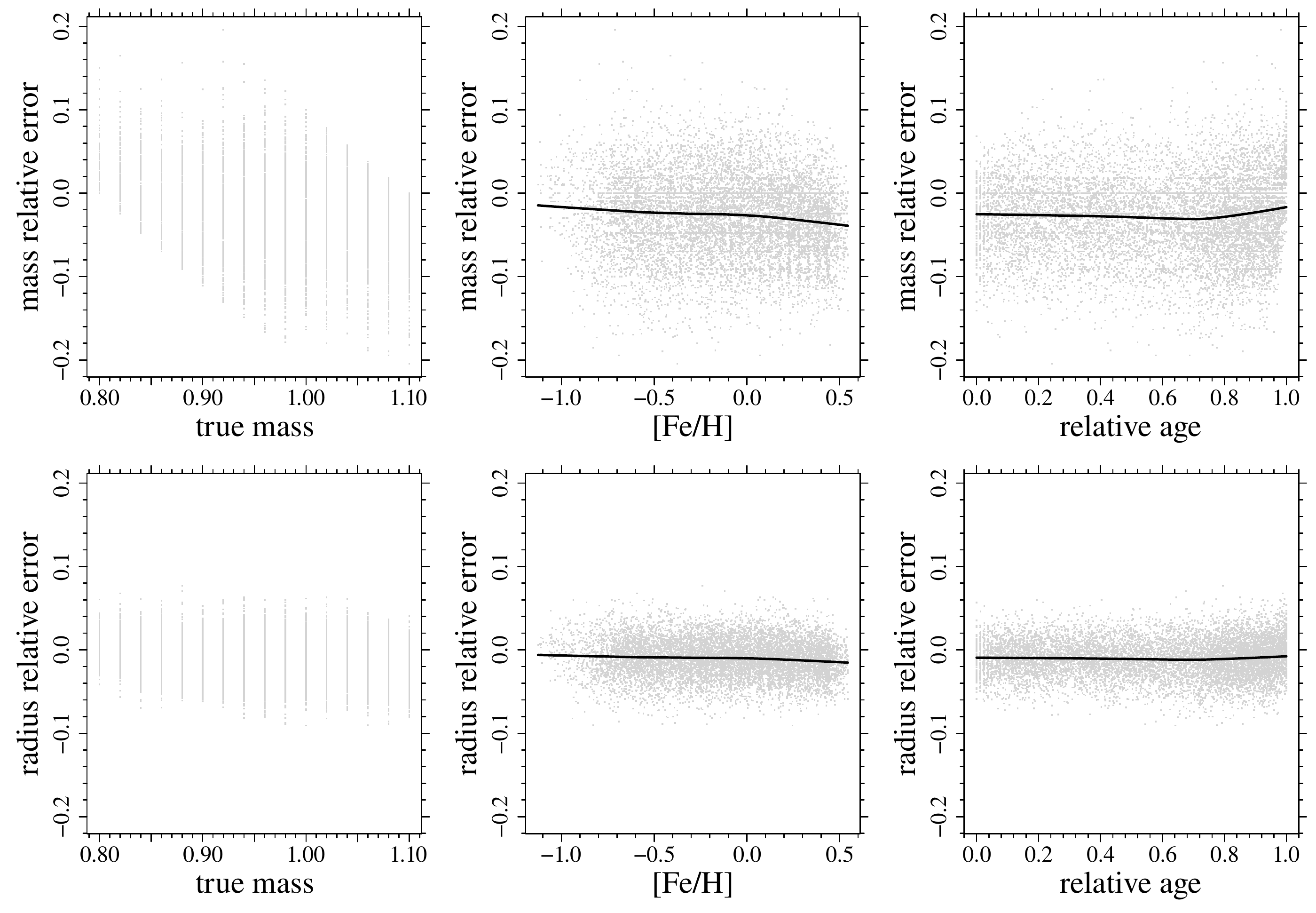}
\caption{
As in Fig.~\ref{fig:std-grid-analisi}, but for synthetic data sampled from a
grid with $\alpha_{\rm ml} = 1.50$, and estimated with the standard grid
(i.e. $\alpha_{\rm ml} = 1.74$).}
\label{fig:ML150-grid-analisi}
\end{figure*}

\begin{figure*}
\centering
\includegraphics[width=11cm,angle=-90]{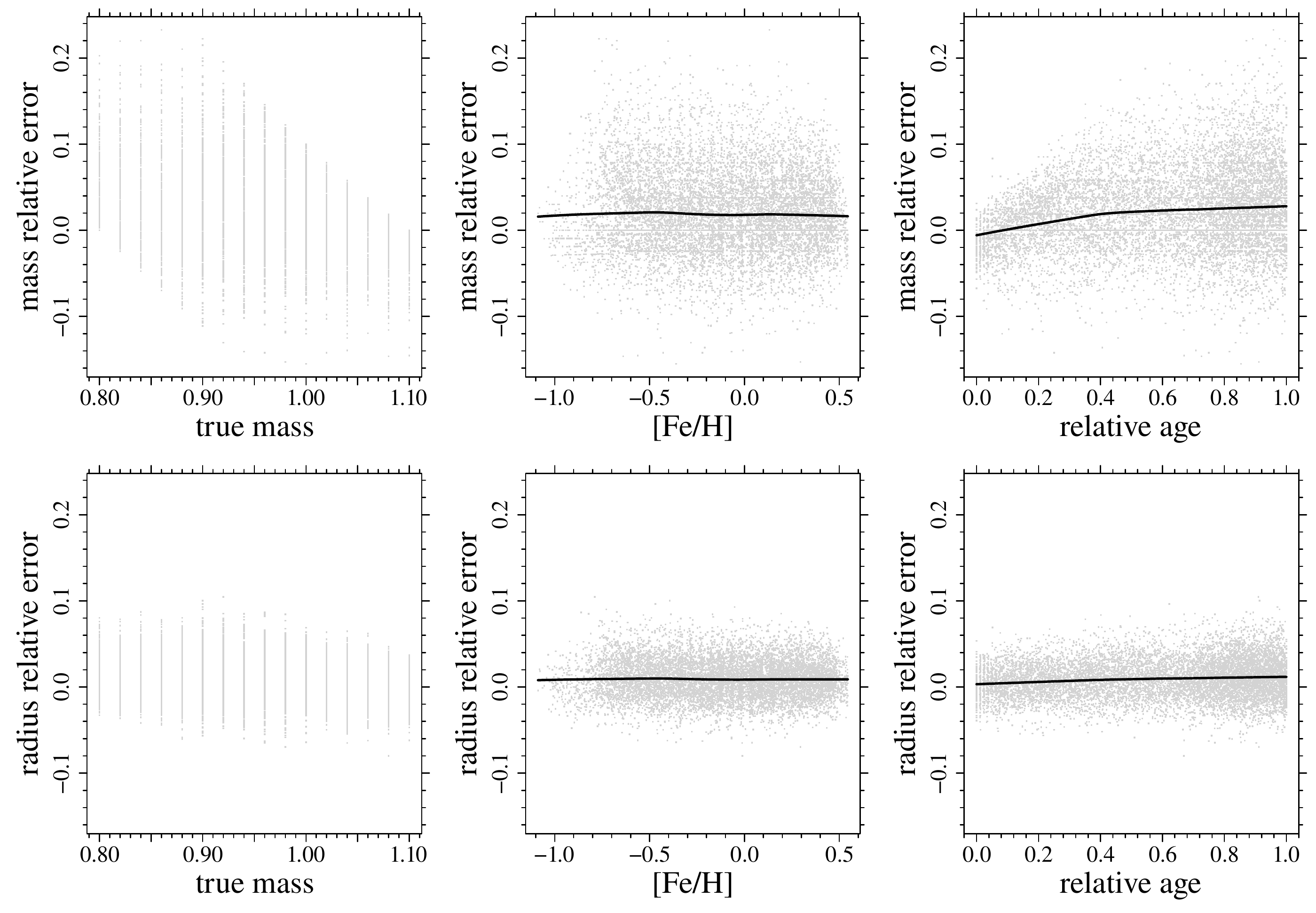}
\caption{
As in Fig.~\ref{fig:std-grid-analisi}, but for synthetic data sampled from a
grid with $\alpha_{\rm ml} = 1.98$ and estimated with the standard grid
(i.e. $\alpha_{\rm ml} = 1.74$).}
\label{fig:ML198-grid-analisi}
\end{figure*}

Figures~\ref{fig:ML150-grid-analisi} and \ref{fig:ML198-grid-analisi} display,
for $\alpha_{\rm ml}$ = 1.50 and 1.98 respectively, the same quantities of
Fig.~\ref{fig:std-grid-analisi}. A comparison of the two figures shows the
different trend in the mass relative error vs. relative age plots.  In detail,
for $\alpha_{\rm ml} = 1.98$, the spread of the relative errors in the
estimated mass is lower at low values of relative ages. The mass is generally
overestimated at high relative ages. For $\alpha_{\rm ml} =
1.50$ instead, the variance of the relative errors is almost constant, and the
mass is underestimated; in the last 20\% of the relative age, the
underestimation is smaller than in the first part of the plot.  All these
trends are edge effects similar to the ones described in
Sect.~\ref{sec:griglia},
and they are the responsible for the deviation from
the linear relation visible in Fig.~\ref{fig:median-ML}. The importance
  of this distortion is greatest for  $\alpha_{\rm ml} = 1.98$. Restricted to
  relative ages greater than 0.35, hence avoiding the strong edge effect
  present in this case, the median of the mass relative error increases from
  1.79\% (Row 7 in Table~\ref{tab:results}) to 2.13\%, which is nearly
    symmetrical to the 
  value $-2.44\%$ obtained 
  for $\alpha_{\rm ml} = 1.50$ (Row 4 in Table~\ref{tab:results}).

The effect of sampling from grids computed with different values of
mixing-length has been previously investigated in \citet{Basu2012}, but with
a different approach. In that work the synthetic grid consisted on a pool of
nine sub-grids with $\alpha_{\rm ml}$ from 1.5 to 2.4 (their solar calibrated
value being 1.826). In that paper three different sets of uncertainty on the
observational quantities are considered; we report here the results for the
Error 2 set (in their Table~1), i.e.  1.0\% in $\Delta \nu$, 2.5\% in
$\nu_{\rm max}$, 100 K in $T_{\rm eff}$, and 0.25 dex in [Fe/H]. The bias
resulting from the reconstruction of the sampled objects -- in the seismic
case -- is 3.50\% in mass and 1.32\% in radius with spreads of
6.86\% and 2.64\%,  respectively. Since the sampling procedure is different, these results
cannot be directly compared with ours. In fact, the results presented in Rows
4-7 of Table~\ref{tab:results} are computed by adopting a unique value of
$\alpha_{\rm ml}$ in the synthetic grid. It is expected that the sampling from
a pool of grids with $\alpha_{\rm ml}$ symmetric around the solar-scaled value
result in negligible bias, except for a small distortion due to edge
effects. The variance is expected to be inflated since it will also account
for the difference in the mean of the estimates obtained from the different
grids in the pool.

\subsection{Radiative opacity, $^{14}$N$(p,\gamma)^{15}$O rate, and diffusion
  velocity} 
\label{sec:physics}

Even in the ideal case in which the metallicity, the helium abundance, and the
mixing-length parameter were exactly known, the resulting stellar models would
still be affected by a non-negligible uncertainty owing to the current degree of
knowledge of the input physics required to solve stellar evolution
equations. In \citet{incertezze1,incertezze2}, we devoted a strong
computational effort to try to quantify the cumulative uncertainty affecting
stellar models due to the combined effect of the main input physics (radiative
opacity, nuclear reaction cross sections, etc.), focussing in particular on
some relevant evolutionary features.

However, the effect of these uncertainties on the stellar parameters estimated
by means of grid-based techniques has never been investigated. This section
represents the first attempt to fill such a gap.

To quantify the impact of the current uncertainties on the
microphysics, we computed additional grids of models by adopting perturbed
input physics. Relying on the results of our previous papers, we can limit
the analysis only to the three input shown to be relevant in the
evolutionary stages and mass range studied here, i.e. the radiative opacities,
the $^{14}$N$(p,\gamma)^{15}$O reaction rate, and the microscopic diffusion
velocities. In light of the results presented in \citet{incertezze1,
  incertezze2}, the impact of the radiative opacity is expected to dominate
the others. As discussed in detail in that paper, we assume an uncertainty of
5\% in radiative opacities, of 15\% in microscopic diffusion velocities, and
of 10\% in the $^{14}$N$(p,\gamma)^{15}$O reaction rate. As shown in our previous
work, as far as perturbations of the input physics of these order are
considered, the interactions among them can be neglected. Therefore, we
computed grids by changing one input at the time while keeping all the others fixed to the
standard values.

In more detail, we computed two grids assuming, respectively, high and low
values of radiative opacities (i.e. $k_{\rm r} \pm $ 5\%), two grids with high
and low values of microscopic diffusion velocities (i.e. $v_{\rm d} \pm $
15\%), and two with high and low values of the $^{14}$N$(p,\gamma)^{15}$O reaction
rate (i.e. $\sigma \pm $ 10\%).

\begin{figure}
\centering
\includegraphics[height=8cm,angle=-90]{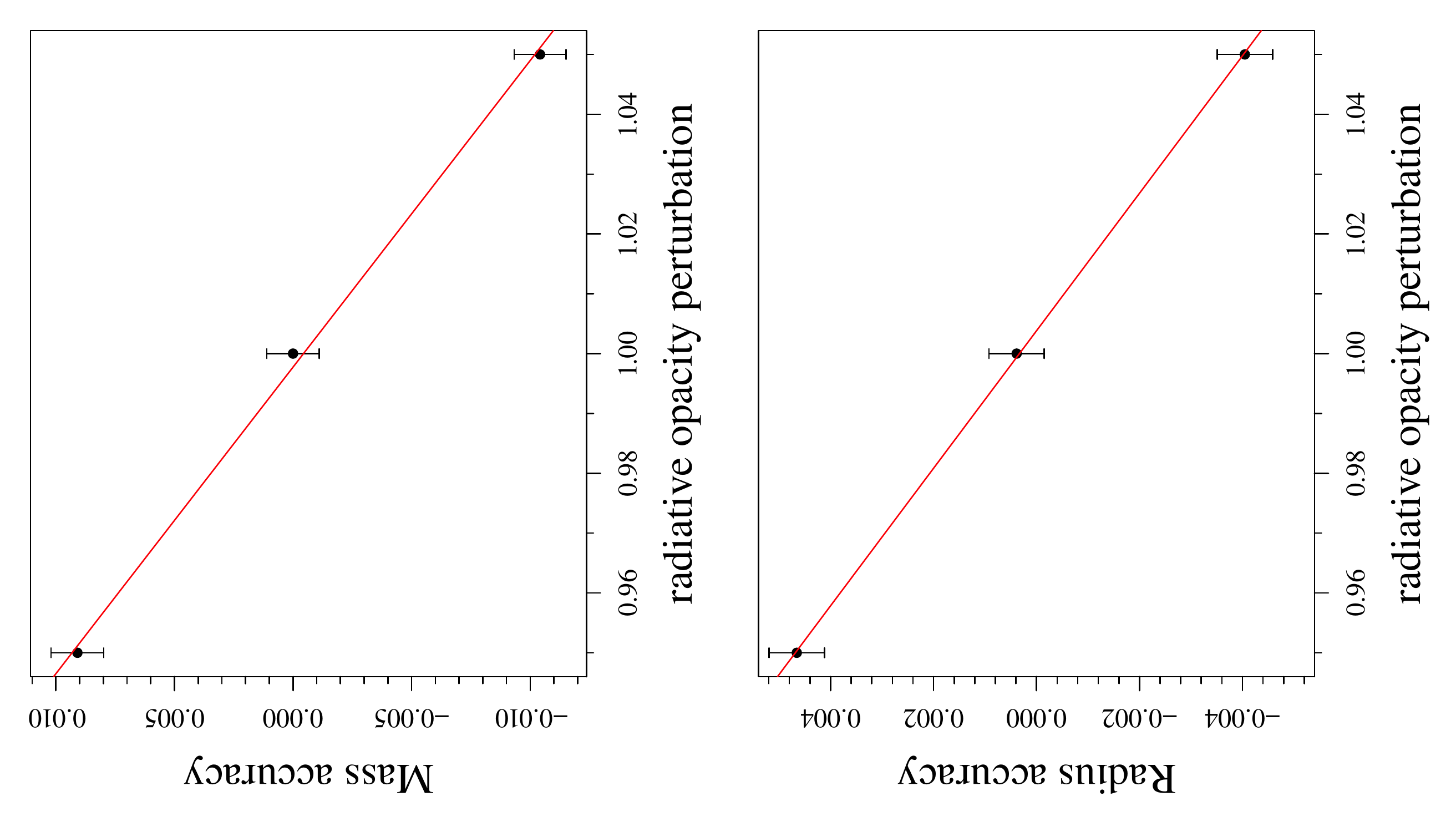}
\caption{
As in Fig.~\ref{fig:median-He}, but 
for synthetic data sampled from grids with different values of radiative
opacity and reconstructed with the standard grid.}
\label{fig:median-kr}
\end{figure}

To assess the importance of these uncertainties, we built synthetic datasets
of artificial stars by sampling $N = 10000$ objects from each of these six
non-standard grids. Then we applied our recovery procedure based on the
standard grid of stellar models on these datasets in order to estimate the
masses and radii. The comparison between recovered and true values allows the
effect of the current uncertainty  in the input physics
adopted in 
stellar evolution codes to be quantified. The results are summarized in the Rows 8-13 of
Table~\ref{tab:results}. Fig.~\ref{fig:median-kr} displays the same quantities
of Fig.~\ref{fig:std-grid-analisi} for the only relevant case, the one with
$k_{\rm r} \pm $ 5\%. In fact, only the radiative opacity perturbation
produces a relevant effect on the mass estimates (see
e.g. Fig.~\ref{fig:median-kr}). For radius estimates the microscopic
diffusion velocities perturbation has an effect of about one half of the one from the opacity
variation. In all the cases the statistical errors dominate over the bias.  

The recovery procedure applied to the two synthetic datasets obtained by adopting
perturbed values of radiative opacity ($k_{\rm r} \pm 5\%$) produces a bias in mass relative error
of about $\mp 1.0\%$ and of about $\mp 0.45\%$ in radius relative errors. As shown in
Fig.~\ref{fig:median-kr}, the effects are symmetric with respect to the
standard value, which corresponds to the case discussed in
Sect.~\ref{sec:results}. These effects are about one half of those
due to 
the initial helium abundance uncertainty and the extreme mixing-length variation
analysed in this work. The trend of the median of the relative errors
distribution with the opacity perturbation can be understood by considering that a lower value of radiative opacity will result in hotter
artificial stars, which mimic more massive stellar models computed with
standard opacity.

\begin{figure*}
\centering
\includegraphics[width=11cm,angle=-90]{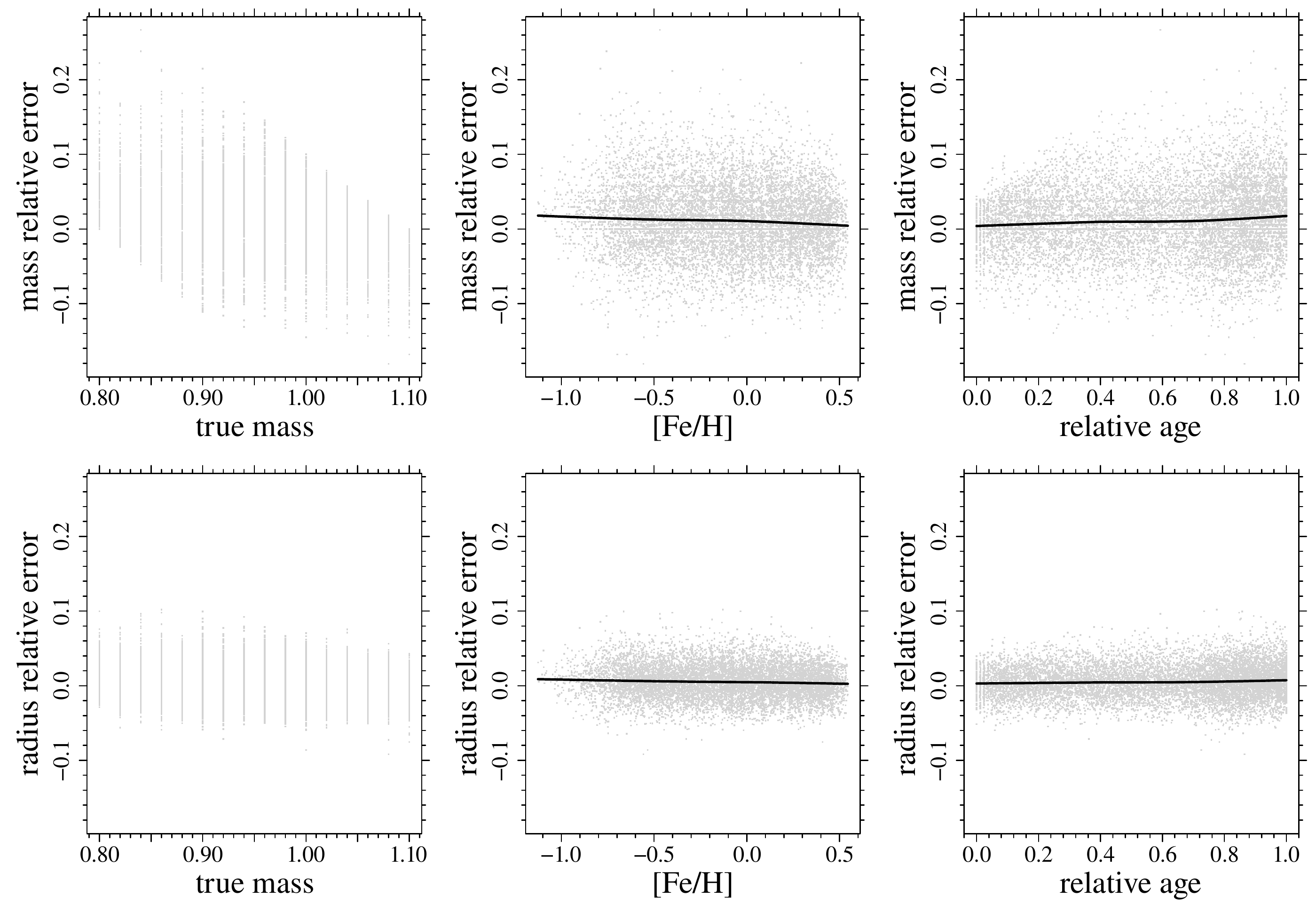}
\caption{
As in Fig.~\ref{fig:std-grid-analisi}, but for synthetic 
data sampled from  a grid with $k_{\rm r}$ at its low value,
estimated on the standard grid.}
\label{fig:kr-grid-analisi}
\end{figure*}

Figure~\ref{fig:kr-grid-analisi} displays the dependence of relative errors on
the mass and radius on the true mass of the stars, on [Fe/H], and on the
relative age of the star. We observe that the performance of the grid-based
estimates does not depend on the metallicity, and the dependence on stellar
  relative age shows a signature of the edge effect, as in the cases discussed
  above.

\subsection{Cumulative effect of mixing-length, helium abundance, and radiative
  opacity uncertainties}  
\label{sec:ml-he-kr}

The possible interactions between simultaneous variations in the mixing-length
parameter and initial helium content or radiative opacity can, in principle,
result in biases in mass and radius estimates that are not the simple sum of
the biases due to the single input.

To test this hypothesis, we computed four further non-standard grids of
stellar models by simultaneously varying two different inputs in order to
explore possible interactions. The first two grids are calculated by combining
the variations in initial helium content and mixing-length parameter: the
former with $\Delta Y/\Delta Z = 1$ and $\alpha_{\rm ml} = 1.50$, the latter
with $\Delta Y/\Delta Z = 3$ and $\alpha_{\rm ml} = 1.98$. The other two
grids take the variation in mixing-length and
radiative opacities simultaneously into account: the former adopting $\alpha_{\rm ml} = 1.98$ and low
values of radiative opacities, the latter with $\alpha_{\rm ml} = 1.50$ and
high values of radiative opacities. These cases correspond to the maximum
variation expected by crossing the selected inputs. Then, we built synthetic
datasets of artificial stars by sampling $N = 10000$ objects from each of
these four non-standard grids of stellar models. The mass and radius of the
objects are finally estimated using the recovery procedure based on the
standard grid.

The results are summarized in Rows 14-17 of Table~\ref{tab:results}.  In
all four cases, the bias of mass and radius estimates are
consistent with the sum of the single biases illustrated in the previous
sections.  In these cases the statistical errors are nearly the same as the
systematic ones. This implies that the cumulative uncertainty in the model
calculations can result in significantly distorted estimates of stellar
characteristics.

\section{Effects of neglecting element diffusion in the stellar model grid} 
\label{sec:feh}

The surface [Fe/H] value observed in real low-mass MS stars clearly represents
the abundance currently present in the atmosphere. Such an abundance can be
quite different from the initial one, depending on the stellar age and mass,
owing to the microscopic diffusion processes. 

\begin{figure}
\centering
\includegraphics[width=8cm,angle=-90]{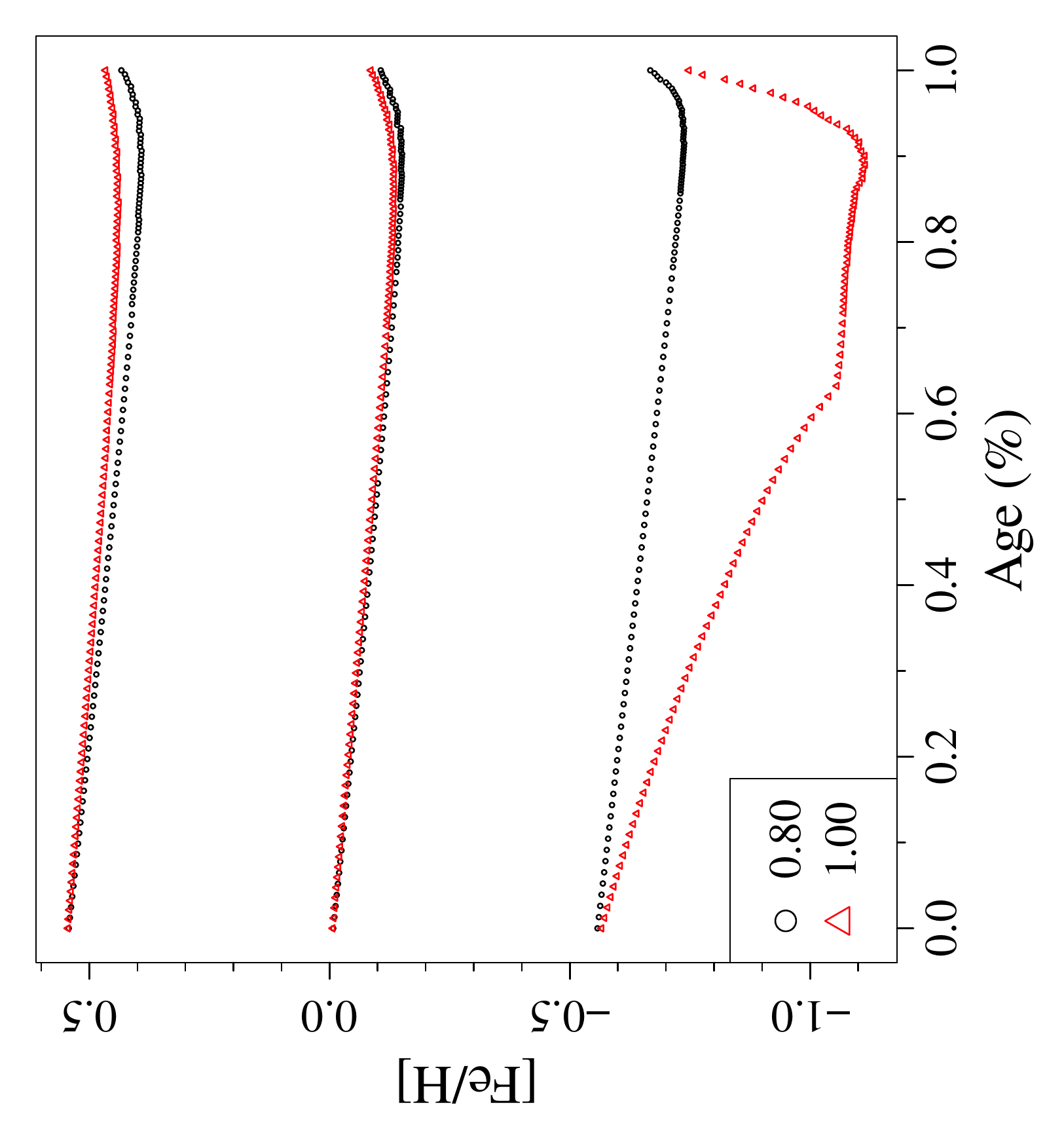}
\caption{Evolution of surface [Fe/H] for two different masses $M$~=~0.80 and 1.00
$M_{\sun}$ (identified by the black circle and red triangle, respectively) and 
 initial [Fe/H] = $-0.55$, 0.00, and 0.55. } 
\label{fig:feh}
\end{figure}

Figure~\ref{fig:feh} shows the [Fe/H] evolution for two stars of $M$ = 0.8 and
1.00 
$M_{\sun}$ with three different initial metallicities ([Fe/H] = $-0.55$, 0.00,
0.55).
The general trend is that, during
the central hydrogen burning, the surface [Fe/H] drops from the ZAMS value and
reaches a minimum at about 90\% of its evolution before central hydrogen
exhaustion. After this point it starts to increase again because of the sink of
the convective envelope, which reaches more internal regions where metals were
previously accumulated by gravitational settling. The amount of the decrease
depends on the mass and initial metallicity  of the star. The lower
the initial metallicity, the higher the drop due to the reduced extension
of the external convective region, which inhibits diffusion. 
The surface [Fe/H] decrease is around 0.1-0.15 dex, 
with the exception of the 1.0 $M_{\sun}$ model with initial [Fe/H] = $-0.55,$  
which shows a dramatic iron depletion (i.e. about 0.55 dex) due to 
the vanishing convective envelope.

These effects of the microscopic diffusion must be taken into account whenever
the stellar parameters of low-mass MS stars have to be determined, since
 neglecting them and using the initial [Fe/H] in
the estimation grid would introduce a systematic bias.  

Nevertheless, some
widely used stellar model grids in the literature, namely BaSTI
\citep{teramo04,teramo06} and STEV \citep{padova08,padova09} do not implement
diffusion. The same occurs in stellar models used in some grid-based
technique, such as RADIUS \citep{Stello2009} and SEEK, which both adopt a grid of
models computed with the Aarhus STellar Evolution Code \citep{Dalsgaard2008}.

Thus, it is useful to analyse the distortion in grid-based
estimates that arises whenever the effects of the diffusion are neglected.
To do this, we follow a different approach to the cases
discussed above, where the recovery procedure has always been performed by means of
the same standard estimation grid of models, and  the various sources of
uncertainty were taken into account in the synthetic dataset
construction. In contrast,  we think that is more realistic in this case to
build the artificial stars by sampling from the standard grid of models, which
takes the element diffusion into account, as in real stars, and to use
non-standard models for the recovery.

Since diffusion affects not only the surface chemical abundance but also the
internal structure and evolution of low-mass MS stars, we study two different
cases. In the first one, the estimation grid stellar models have been computed
without element diffusion, whereas in the second one diffusion is allowed, but
the surface evolution of [Fe/H] is neglected in the recovery procedure
(i.e. the [Fe/H] value in the evolutionary tracks is fixed to be equal to its initial
value). In the latter case, the effect on the reconstructed mass and radius is
only due to an incorrect initial metallicity evaluation, since one would assign an initial metal abundance equal to the observed surface
one
to the observed star, thus underestimating the metallicity. In the former, one would
also neglect the other evolutionary effects of diffusion.

\begin{figure*}
\centering
\includegraphics[width=11cm,angle=-90]{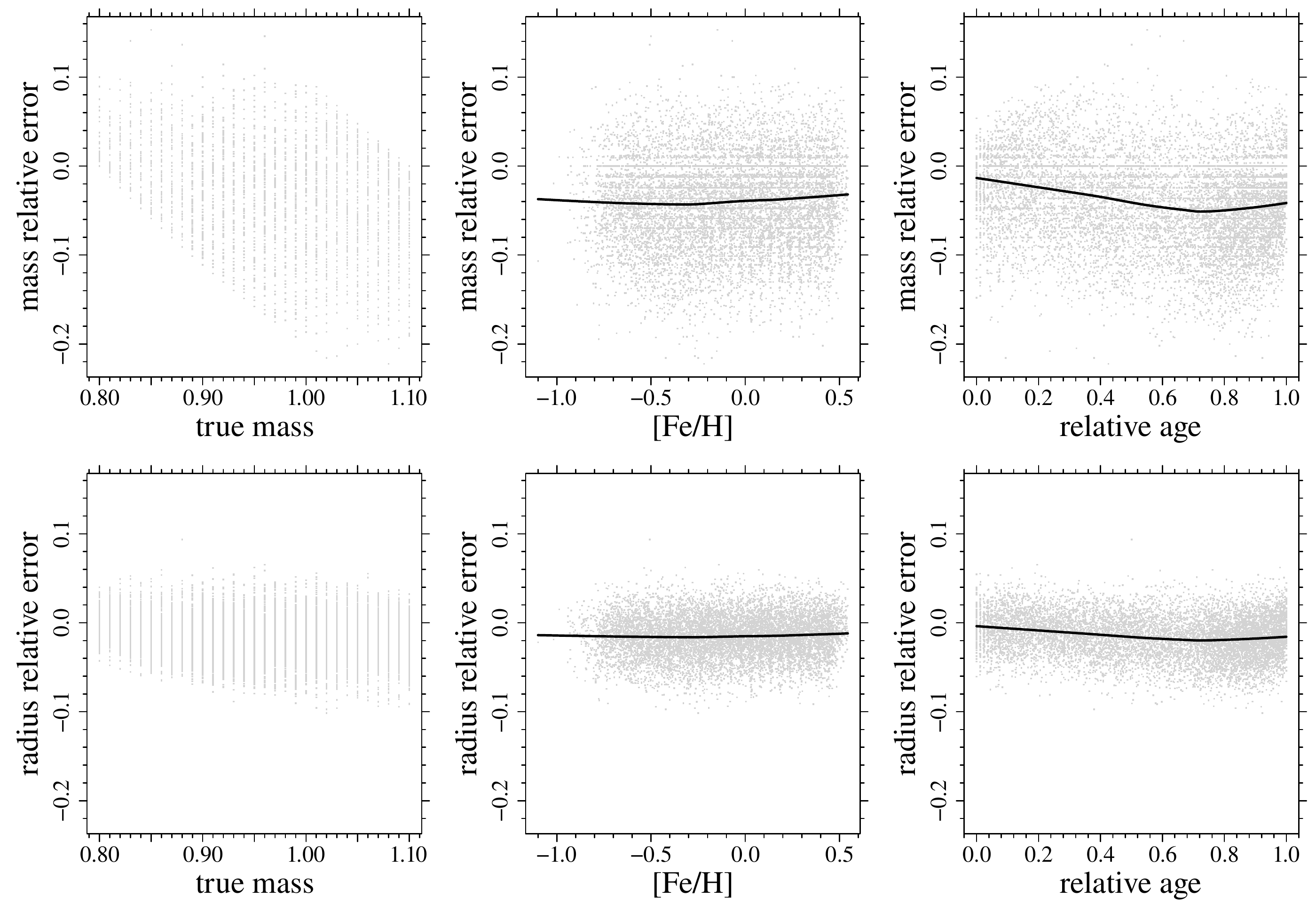}
\caption{
As in Fig.~\ref{fig:std-grid-analisi}, but for synthetic 
data sampled from the standard grid and recovered on the
grid, which does not include diffusion in the computations.}
\label{fig:nodiff-grid-analisi}
\end{figure*}

The results of the first case, i.e. completely neglecting the diffusion in
the estimation grid, are summarized in Rows 18 of Table~\ref{tab:results}. The
bias in mass relative error is about $-3.7\%$, while it is about $-1.5\%$ for
radius relative errors. The statistical errors, about $5.0\%$ and $2.3\%$
respectively, are very close to the biases. The biases turn out to be
comparable to the one presented in Row 17 of Table~\ref{tab:results}, obtained
for the combined variation of mixing-length parameter and radiative
opacity. Figure~\ref{fig:nodiff-grid-analisi} shows the dependence of relative
errors in the mass and radius on the true mass of the stars, on [Fe/H], and on
the relative age of the star. As expected, the effect of microscopic diffusion
is greater at higher relative ages, since the timescale of the microscopic
diffusion is a few Gyr. As a consequence the bias in the mass and
radius estimates is greater for
stars in late central hydrogen burning. Figure~\ref{fig:nodiff-grid-analisi}
also shows that around [Fe/H]~=~$-0.5,$ the underestimation of mass and radius is
more than for metal-richer stars. This effect is the due to the
metallicity dependence of the 
external convection extension, which inhibits diffusion, hence of
the diffusion efficiency.

The results of the second case are summarized in Row 19 of
Table~\ref{tab:results}.  The bias in mass relative error is about $-2.6\%$,
while it is about $-1\%$ for radius relative errors. The statistical errors,
about $4.8\%$ and $2.3\%$ respectively, are dominant in both cases.
 
\begin{figure*}
\centering
\includegraphics[width=11cm,angle=-90]{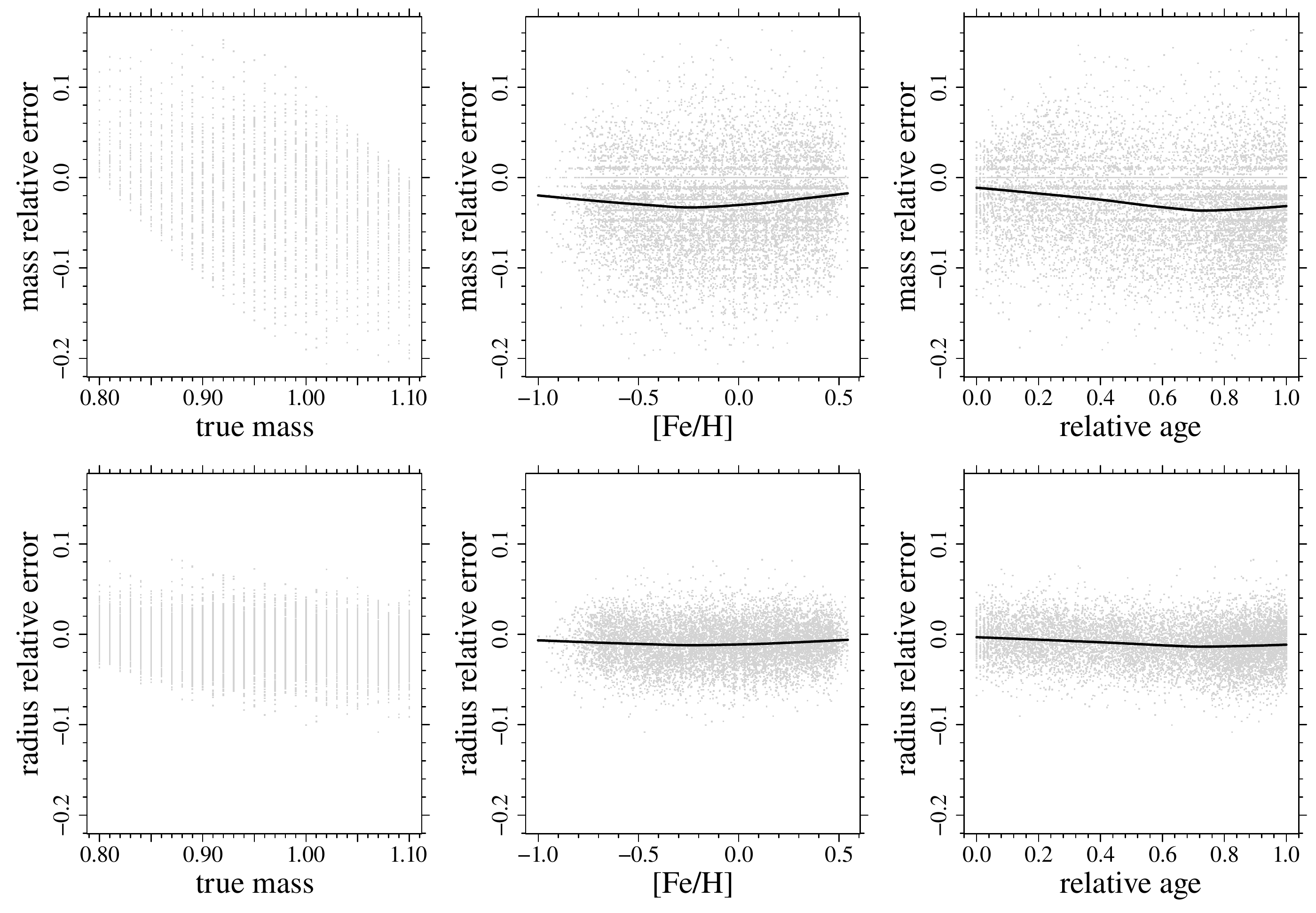}
\caption{
As in Fig.~\ref{fig:std-grid-analisi}, but for synthetic 
data sampled from the standard grid and recovered on the
  same grid but whithout taking the surface [Fe/H] evolution into account.}
\label{fig:feh-grid-analisi}
\end{figure*}

As one can see in Row 2 of Table~\ref{tab:results}, these values are nearly
the same as those due to the initial helium abundance uncertainty. From
Fig.~\ref{fig:feh-grid-analisi}, it appears that the trend of the relative
errors in mass and radius are similar to those discussed above that result from
the complete neglecting of diffusion. 

The comparison of the results reported in Rows 18 and 19 in
Table~\ref{tab:results} shows that the neglecting of surface [Fe/H] evolution
is the main bias source, since it accounts for about two-thirds of the overall bias
due to completely neglecting diffusion in stellar evolution.

\section{Possible extensions of the standard estimation grid} 
\label{sec:extension}

An extension of the estimation grid with models computed with different 
input physics and parameters might in principle improve mass and 
radius estimates in the presence of unknown sources of uncertainty. 
The general idea is that the presence
of grids computed with different mixing-length values or different radiative
opacities can mimic the variation induced by other sources of uncertainty,
e.g. the initial helium contents. As a result the additional grids would help
to keep the hidden variability under control, providing less biased estimates.
A similar method has already been adopted in the literature \citep[see
  e.g.][]{Basu2012}, although with some differences to the
computations presented here.

To check such a working hypothesis, we built synthetic datasets of artificial
stars by sampling from a grid of models not included in the estimation
process.  In other words, we did not sample the synthetic stars from the same
grids used in the recovery procedure, unlike \citet{Basu2012}. This
is a subtle but important distinction. If the synthetic dataset were built
from the same extended grid of models as used for the subsequent recovery
procedure, the resulting estimates would by necessity be less biased than the
ones obtained from the standard grid alone, since in this case the extended
grid indeed also contains the sampled synthetic stars. This is clearly not
the case when synthetic stars are sampled from grids that are  not included.

To assess the usefulness of the grid extension we adopted the following
approach. As a first test, we built two synthetic datasets of artificial stars
by sampling $N = 10000$ objects from non-standard grids of stellar models with
mixing-length values different from the solar calibrated one, namely 
$\alpha_{\rm ml}$ = 1.50 and 1.98. Then we applied the recovery procedure to
estimating the mass and radius of these objects and added all the stellar models computed with different values of
initial helium abundance to our standard
estimation grid (i.e.  $\Delta
Y/\Delta Z$ = 1 and 3).  The results are summarized in Rows 1 and 2 of
Table~\ref{tab:results-nostsndard}.

As a second test, we built two other synthetic datasets of artificial stars by
sampling $N = 10000$ objects from the non-standard grids of stellar models
computed with high and low radiative opacities values. Then we reconstructed
the mass and radius by adopting the same
extended grid used in the previous test in the recovery procedure (results in Rows 3-4 of
Table~\ref{tab:results-nostsndard}).
   
Finally, we analysed the same synthetic dataset as for the first
test (i.e. non-solar mixing-length parameters) but adopted a
different extended grid that includes the two non-standard grids computed
with high and low radiative opacities beyond the standard one (results in Rows
5-6 of Table~\ref{tab:results-nostsndard}).

The comparison of the results reported in Table~\ref{tab:results-nostsndard}
(Rows 1-4) with those presented in Table~\ref{tab:results} (Rows 4, 7, 8, 9)
shows that the insertion of non-standard grids computed with different $\Delta
Y/\Delta Z$ values produces estimates of mass and radius with a slightly reduced bias,
but at the expense of an increase in the statistical error by about 15\%. The
estimates obtained from the grid that includes non-standard models with
different values of the radiative opacity (Rows 5-6 in
Table~\ref{tab:results-nostsndard}) do not differ significantly from the ones
obtained with the standard grid only (Rows 4 and 7 in Table~\ref{tab:results}).

The importance of a multi-mixing-length recovery grid can be different for
different stellar evolutionary stages or for different sources of uncertainty.  
As an example, in \cite{Basu2012} -- which considers stars up to 3.0 $M_{\sun}$,
including the red giant phase --
it was stated that, for stars sampled from a grid computed with
different boundary conditions, the bias of the mass
estimates is lower in the multi-grid case.

\begin{table*}[ht]
\centering
\caption{Summary of mass and radius relative errors.}
\label{tab:results}
\begin{tabular}{rlrrrrrr}
  \hline\hline
 & Label & Median & 95\% CI & Std. dev. & 95\% CI & $q_{16}$ & $q_{84}$ \\ 
  \hline
\multicolumn{8}{c}{Mass estimate}\\
 \hline
  1 & standard &  0.0000 & 0.0011 & 0.0450 & 0.0006 & -0.043 & 0.042 \\ 
  2 & $\Delta Y/\Delta Z$ = 1 & -0.0244 & 0.0012 & 0.0487 & 0.0007 & -0.075 & 0.017 \\ 
  3 & $\Delta Y/\Delta Z$ = 3 &  0.0227 & 0.0012 & 0.0490 & 0.0007 & -0.019 & 0.075 \\ 
  4 & $\alpha_{\rm ml}$ = 1.50 & -0.0244 & 0.0011 & 0.0468 & 0.0006 & -0.074 & 0.016 \\ 
  5 & $\alpha_{\rm ml}$ = 1.62 & -0.0104 & 0.0011 & 0.0454 & 0.0006 & -0.058 & 0.029 \\ 
  6 & $\alpha_{\rm ml}$ = 1.86 &  0.0079 & 0.0011 & 0.0455 & 0.0006 & -0.031 & 0.054 \\ 
  7 & $\alpha_{\rm ml}$ = 1.98 &  0.0179 & 0.0011 & 0.0463 & 0.0006 & -0.020 & 0.067 \\ 
  8 & $k_{\rm r}$ low &  0.0091 & 0.0011 & 0.0453 & 0.0006 & -0.030 & 0.055 \\ 
  9 & $k_{\rm r}$ high & -0.0104 & 0.0011 & 0.0446 & 0.0006 & -0.055 & 0.028 \\ 
  10 & $^{14}$N$(p,\gamma)^{15}$O low &  0.0000 & 0.0011 & 0.0451 & 0.0006 & -0.044 & 0.041 \\ 
  11 & $^{14}$N$(p,\gamma)^{15}$O high &  0.0000 & 0.0011 & 0.0444 & 0.0006 & -0.044 & 0.040 \\ 
  12 & $v_{\rm d}$ low &  0.0000 & 0.0011 & 0.0454 & 0.0006 & -0.038 & 0.048 \\ 
  13 & $v_{\rm d}$ high & -0.0027 & 0.0011 & 0.0447 & 0.0006 & -0.048 & 0.038 \\ 
  14 & $\alpha_{\rm ml}$ = 1.50, $\Delta Y/\Delta Z$ = 1 & -0.0476 & 0.0013 & 0.0515 & 0.0007 & -0.104 & 0.000 \\ 
  15 & $\alpha_{\rm ml}$ = 1.98, $\Delta Y/\Delta Z$ = 3 &  0.0395 & 0.0013 & 0.0514 & 0.0007 &  0.000 & 0.100 \\ 
  16 & $\alpha_{\rm ml}$ = 1.98, $k_{\rm r}$ low &  0.0284 & 0.0011 & 0.0463 & 0.0006 & -0.009 & 0.078 \\ 
  17 & $\alpha_{\rm ml}$ = 1.50, $k_{\rm r}$ high & -0.0365 & 0.0011 & 0.0465 & 0.0006 & -0.085 & 0.003 \\ 
  18 & no diffusion & -0.0370 & 0.0012 & 0.0496 & 0.0007 & -0.091 & 0.008 \\
  19 & no [Fe/H] evolution & -0.0257 & 0.0012 & 0.0481 & 0.0007 & -0.076 & 0.014 \\ 
  \hline
\multicolumn{8}{c}{Radius estimate}\\
 \hline
  1 & standard &  0.0004 & 0.0005 & 0.0219 & 0.0003 & -0.021 & 0.022 \\ 
  2 & $\Delta Y/\Delta Z$ = 1 & -0.0106 & 0.0006 & 0.0235 & 0.0003 & -0.033 & 0.013 \\ 
  3 & $\Delta Y/\Delta Z$ = 3 &  0.0109 & 0.0006 & 0.0233 & 0.0003 & -0.012 & 0.034 \\ 
  4 & $\alpha_{\rm ml}$ = 1.50 & -0.0103 & 0.0005 & 0.0222 & 0.0003 & -0.032 & 0.012 \\ 
  5 & $\alpha_{\rm ml}$ = 1.62 & -0.0043 & 0.0005 & 0.0219 & 0.0003 & -0.026 & 0.017 \\ 
  6 & $\alpha_{\rm ml}$ = 1.86 &  0.0049 & 0.0005 & 0.0219 & 0.0003 & -0.016 & 0.027 \\ 
  7 & $\alpha_{\rm ml}$ = 1.98 &  0.0088 & 0.0005 & 0.0219 & 0.0003 & -0.012 & 0.031 \\ 
  8 & $k_{\rm r}$ low &  0.0047 & 0.0005 & 0.0219 & 0.0003 & -0.016 & 0.027 \\ 
  9 & $k_{\rm r}$ high & -0.0040 & 0.0005 & 0.0219 & 0.0003 & -0.025 & 0.017 \\ 
  10 & $^{14}$N$(p,\gamma)^{15}$O low &  0.0003 & 0.0005 & 0.0223 & 0.0003 & -0.021 & 0.022 \\ 
  11 & $^{14}$N$(p,\gamma)^{15}$O high & -0.0002 & 0.0005 & 0.0215 & 0.0003 & -0.020 & 0.022 \\ 
  12 & $v_{\rm d}$ low &  0.0021 & 0.0005 & 0.0218 & 0.0003 & -0.018 & 0.024 \\ 
  13 & $v_{\rm d}$ high & -0.0015 & 0.0005 & 0.0218 & 0.0003 & -0.022 & 0.020 \\ 
  14 & $\alpha_{\rm ml}$ = 1.50, $\Delta Y/\Delta Z$ = 1 & -0.0199 & 0.0006 & 0.0244 & 0.0003 & -0.044 & 0.004 \\ 
  15 & $\alpha_{\rm ml}$ = 1.98, $\Delta Y/\Delta Z$ = 3 &  0.0187 & 0.0006 & 0.0239 & 0.0003 & -0.004 & 0.044 \\ 
  16 & $\alpha_{\rm ml}$ = 1.98, $k_{\rm r}$ low &  0.0126 & 0.0005 & 0.0220 & 0.0003 & -0.008 & 0.035 \\ 
  17 & $\alpha_{\rm ml}$ = 1.50, $k_{\rm r}$ high & -0.0144 & 0.0005 & 0.0224 & 0.0003 & -0.037 & 0.007 \\ 
  18 & no diffusion & -0.0149 & 0.0006 & 0.0234 & 0.0003 & -0.038 & 0.008 \\ 
  19 & no [Fe/H] evolution & -0.0104 & 0.0006 & 0.0228 & 0.0003 & -0.033 & 0.012 \\ 
    \hline
\end{tabular}
\tablefoot{In the first column: grid employed for sampling; in the second and third
  columns: median and width of its 95\% confidence interval for the relative
  errors of mass/radius estimates; in the fourth and fifth columns: standard
  deviation and width of its 95\% confidence interval for the relative 
  errors of mass/radius estimates; sixth and seventh columns: 16th and 84th
  quantiles for the relative 
  errors of mass/radius estimates.}
\end{table*}

\begin{table}[ht]
\centering
\caption{Summary of standard deviation of mass and radius relative errors
  obtained adopting reduced  errors on observation constraints. For the observables not 
  explicitly indicated in the labels, the errors are the same as those adopted in the standard case.}
\label{tab:results-rederror}
\begin{tabular}{rlrrr}
  \hline\hline
 & Label & std. dev. & $q_{16}$ & $q_{84}$ \\ 
\hline
\multicolumn{5}{c}{Mass estimate}\\
 \hline
1 & $\sigma(T_{\rm eff}) =$ 50 K & 0.0350 & -0.034 & 0.031 \\
2 & $\sigma([Fe/H]) =$ 0.05 dex      & 0.0392 & -0.038 & 0.037 \\
3 & $\sigma(\Delta \nu) =$ 1\%, $\sigma(\nu_{\rm max}) =$ 2.5\% & 0.0427 & -0.041 & 0.040 \\  
4 & all the above            & 0.0246 & -0.023 & 0.023 \\ 
\hline
\multicolumn{5}{c}{Radius estimate}\\
 \hline
1 & $\sigma(T_{\rm eff}) =$ 50 K & 0.0193 & -0.018 & 0.020 \\
2 & $\sigma([Fe/H]) =$ 0.05 dex      & 0.0203 & -0.019 & 0.020 \\
3 & $\sigma(\Delta \nu) =$ 1\%, $\sigma(\nu_{\rm max}) =$ 2.5\%    & 0.0164 & -0.016 & 0.016 \\   
4 & all the above            & 0.0108 & -0.010 & 0.011 \\
 \hline
\end{tabular}
\end{table}

\begin{table*}[ht]
\centering
\caption{Summary of mass and radius relative errors obtained adopting non-standard estimation grid. }
\label{tab:results-nostsndard}
\begin{tabular}{rllrrrrrr}
  \hline\hline
 & Label & Estimation grid & Median & 95\% CI & Std. dev. & 95\% CI & $q_{16}$ & $q_{84}$ \\ 
  \hline
\multicolumn{9}{c}{Mass estimate}\\
 \hline
  1 & $\alpha_{\rm ml}$ = 1.50 & std. + He & -0.0208 & 0.0014 & 0.0552 & 0.0008 & -0.080 & 0.028 \\ 
  2 & $\alpha_{\rm ml}$ = 1.98 & std. + He &  0.0111 & 0.0013 & 0.0533 & 0.0007 & -0.032 & 0.070 \\ 
  3 & $k_{\rm r}$ low & std. + He &  0.0062 & 0.0013 & 0.0529 & 0.0007 & -0.041 & 0.060 \\ 
  4 & $k_{\rm r}$ high & std. + He & -0.0096 & 0.0013 & 0.0533 & 0.0007 & -0.062 & 0.038 \\ 
  5 & $\alpha_{\rm ml}$ = 1.50 & std. + $k_{\rm r}$ & -0.0244 & 0.0012 & 0.0470 & 0.0007 & -0.073 & 0.019 \\ 
  6 & $\alpha_{\rm ml}$ = 1.98 & std. + $k_{\rm r}$ &  0.0160 & 0.0012 & 0.0472 & 0.0007 & -0.024 & 0.066 \\ 
  \hline
\multicolumn{9}{c}{Radius estimate}\\
 \hline
  1 & $\alpha_{\rm ml}$ = 1.50 & std. + He & -0.0081 & 0.0006 & 0.0253 & 0.0004 & -0.034 & 0.016 \\ 
  2 & $\alpha_{\rm ml}$ = 1.98 & std. + He &  0.0067 & 0.0006 & 0.0244 & 0.0003 & -0.016 & 0.032 \\ 
  3 & $k_{\rm r}$ low & std. + He &  0.0040 & 0.0006 & 0.0243 & 0.0003 & -0.019 & 0.028 \\ 
  4 & $k_{\rm r}$ high & std. + He & -0.0042 & 0.0006 & 0.0246 & 0.0003 & -0.028 & 0.021 \\ 
  5 & $\alpha_{\rm ml}$ = 1.50 & std. + $k_{\rm r}$ & -0.0098 & 0.0005 & 0.0224 & 0.0003 & -0.032 & 0.012 \\ 
  6 & $\alpha_{\rm ml}$ = 1.98 & std. + $k_{\rm r}$ &  0.0085 & 0.0005 & 0.0224 & 0.0003 & -0.013 & 0.031 \\ 
   \hline
\end{tabular}
\tablefoot{Column labels are the same as in Table~\ref{tab:results}.}
\end{table*}

\section{Comparison with other techniques}
\label{sec:comparison}

Computation of the bias on the estimated mass and radius presented in the
previous sections shed some light on the magnitude of systematic errors due to
some uncertainties affecting stellar evolution. These computations have been
obtained using a single evolutionary code,  so they share a large number of
 input physics and algorithmic approaches. The comparison with the results
obtained by other authors on a common set of objects is therefore of high
interest for estimating the possible contributions to the systematic uncertainty
arising from different stellar evolutionary computations and from different
recovery techniques.

In this section we apply our standard estimation grids to recovering stellar
parameters of seven objects -- K3656476, K6116048, K7976303, K8006161,
K8379927, K10516096, and K10963065 -- from the Kepler catalogue. The selected
objects have been extensively studied by \citet{Mathur2012}, along with other
15 objects that lie outside our estimation grid, by adopting RADIUS, YB, and SEEK.

The RADIUS method \citep{Stello2009} uses $T_{\rm eff}$, $\log g$, [Fe/H],
$L$, and $\Delta \nu$ to find the optimal model. It is based on a large grid
of stellar models computed with the Aarhus STellar Evolution Code
\citep{Dalsgaard2008}, which does not include rotation, overshooting, and
diffusion. The grid has fixed values of the mixing-length and initial helium
abundance at a given metallicity. The \citet{GN93} solar mixture is
adopted. The metallicity $Z$ is in the range [0.001 - 0.055], and the mass step
is 0.01 $M_{\sun}$ in the range [0.5 - 4.0] $M_{\sun}$. The recovery technique
identifies a $3 \sigma$ region in the observation hyperspace around the object
to reconstruct, finds the extreme values of mass and radius in this set, and
reports their averages as mass and radius estimates.

The YB method uses a variant of the Yale-Birmingham code \citep{Gai2011} and
adopts the same estimation technique as in the present work. The recovery is
based on the following observables: $T_{\rm eff}$, [Fe/H], $\Delta \nu$, and
$\nu_{\rm max}$. The grid of stellar models is computed with the Yale rotating
evolution code \citep{Demarque2008} in its non-rotating configuration.  The
grid spans the mass range [0.80 - 3.0] $M_{\sun}$ in steps of 0.02 $M_{\sun}$.
The metallicity [Fe/H] ranges from $-0.6$ to 0.6 dex, with solar abundance
according to \citet{GS98}. The initial helium abundance is linked to the
metallicity assuming $\Delta Y/\Delta Z = 1$.

The SEEK technique requires the use of a large grid of stellar models computed
with the Aarhus Stellar Evolution Code \citep{Dalsgaard2008}, and it adopts a
Bayesian approach in the estimation procedure. The grid is composed of 7300
evolution tracks with different mixing-length parameters and initial helium
abundances at a given metallicity.  It adopts the solar mixture of
\citet{GS98}. The mass step in the range [0.6 - 1.8] $M_{\sun}$ is 0.02
$M_{\sun}$. The metallicity range of the grid is $Z$ $\in$ [0.005, 0.03].
The observables used in the reconstruction are $T_{\rm eff}$, [Fe/H], and $\Delta
\nu$.

Table~\ref{tab:obs-K} lists the observational constraints adopted in the
estimation.  Regarding the seismic quantities, we adopt a slightly
conservative approach -- which accounts for the uncertainty in the solar
seismic values -- and we assume a common uncertainty of 1\% in $\Delta \nu$
and 5\% in $\nu_{\rm max}$.

In Table~\ref{tab:estim-obs} we present the results of the estimation
procedure. The table lists the results obtained using only the standard grid
and those obtained also using the grids of stellar models with
different mixing-length values and different initial helium contents. The
second approach is similar to the one adopted in the SEEK technique. The
corresponding estimates quoted in \citet{Mathur2012} are in
Table~\ref{tab:estim-ref}.  As noted in Sect.~\ref{sec:extension}, the
statistical error from the multi-grid estimation technique is often larger
than the one from the single-grid technique.  
This effect can be anticipated since the estimates in the multi-grid cases 
can be viewed as the pool
of the estimates obtained on the different grids inserted in the recovery
procedure. All these estimates have almost equal variance (see
Table \ref{tab:results}). The total variance of the mass and radius estimates
are then slightly
inflated by the presence of the systematic bias of the estimates on the
different grids in the recovery procedure.
This small amount of inflaction confirms 
that the statistical component of the error term is more important than the
systematic shift in the estimates due to the differences in the
grids. Obviously, an improvement on the precision of the stellar observables
will modify the balance of these two error terms, since the statistical errors
will shrink \citep[see e.g.][]{Gai2011}.

Although the adopted technique is the same as YB, the results of
the present work match those provided by the SEEK technique best. In fact in all
the studied cases, our and SEEK estimates of both mass and radius are consistent
within the errors.
This results illustrates the difficulties in disentangling 
the 
various input and techniques adopted in the grid estimation procedure. 
The SEEK grid does not include microscopic diffusion, which is shown here to
contribute largely to biasing the estimates of mass and radius, but this bias
is cancelled out by the other differences in the evolutionary code and in the
estimation technique.

Comparison with the estimates obtained by YB is highly informative since
it highlights the systematic uncertainty arising only from the differences
in the stellar evolution computations, because the recovery procedure is the same. 
The YB mass estimates of K6116048 and
K8379927 are significantly higher -- at $1 \sigma$ level -- than the ones
obtained here using the standard reconstruction grid, but consistent with the
estimates obtained using all the other grids. As for radii, YB estimates for
K6116048, K7976303, and K8379927, are significantly larger than the ones
obtained here, while for K10516096 the estimate is significantly lower.

A general agreement among the estimation techniques is not unexpected. In
fact, we are focussing on central hydrogen burning stars in a narrow range of
masses around the solar one. In these cases, the common procedure of
calibrating the mixing-length parameter to the Sun will keep
most of the differences induced by the different input physics in the
evolutionary codes under control. A poorer agreement is expected whenever later stages,
e.g. the red giant branch evolution, are considered. Nevertheless, even from
the analysis of the few objects presented here, we note that the differences
in the assumptions made in the different codes (input physics, convection
and diffusion efficiencies, chemical composition, etc.)  play a fundamental
role. The mass estimates provided by all of the techniques agree
within the errors only for K7976303, K8006161, K10516096, and K10963065, while
the agreement of both mass and radius is only obtained for K10963065.

A simple exercise gives an idea of the range of mass spanned by the
  different estimation 
techniques. We averaged the differences in the higher and lower estimates
obtained by the four techniques for all the seven objects. The result is
$\left<\Delta M\right> = 0.14$ $M_{\sun}$, which is much higher than
the statistical components of the error obtained by the grid techniques. 
As a reference, the mean of these latter errors is about 0.05 $M_{\sun}$.
 The
same computation performed on radii gives an average of $\left<\Delta
R\right> = 0.11$ $R_{\sun}$; even in this case, the systematic error is
more than the statistical ones (the mean of statistical errors is about
0.02 $R_{\sun}$). 

It is apparent that the magnitude of the systematic error shown here is
much larger than the one reported in the analyses of the previous sections and
that it is dominant over the statistical component. It appears that the
differences in the evolutionary codes and in the recovery techniques play a
fundamental role in estimating stellar parameters. Since the
refinement of the observation techniques will make data available with
increasing precision, it is expected that this conclusion will strengthen in the
near future, since better observation precision will obviously lead to smaller
statistical errors, as displayed in Table~\ref{tab:results-rederror} \citep[see
  also][]{Gai2011}.

As a final remark, the asteroseismic radius of KIC 8006161
  has been verified using parallaxes \citep{SilvaAguirre2012} and
  interferometry \citep{Huber2012}. The values are  $R = 0.927
  \pm 0.014$ $R_{\sun}$ and $R = 0.952
  \pm 0.021$ $R_{\sun,}$ respectively. The results reported here ($R = 0.92
  \pm 0.01$ $R_{\sun}$ for single grid estimate, and  $R = 0.93
  \pm 0.02$ $R_{\sun}$  for multi-grid estimate) agree with
  these determinations.

\begin{table*}[ht]
\centering
\caption{Observational sample of stars from the Kepler catalog selected for the estimation procedure. }
\label{tab:obs-K}
\begin{tabular}{lcccc}
  \hline\hline
Star & $T_{\rm eff}$ (K)& [Fe/H] & $\Delta \nu$ ($\mu$Hz) & $\nu_{\rm max}$
($\mu$Hz) \\
\hline
K3656476 &  5700 $\pm$ 70 & 0.32 $\pm$ 0.07 & 93.70 $\pm$ 0.22 & 1940 $\pm$ 25 \\
K6116048 &  5895 $\pm$ 70 & -0.26 $\pm$ 0.07 & 100.14 $\pm$ 0.22 & 2120 $\pm$ 20 \\
K7976303 & 6050 $\pm$ 70 & 0.10 $\pm$ 0.07 & 50.95 $\pm$ 0.37 & 910 $\pm$ 25 \\ 
K8006161 & 5340 $\pm$ 70 & 0.38 $\pm$ 0.07 & 148.21 $\pm$ 0.19 & 3545 $\pm$ 140 \\
K8379927 & 5960 $\pm$ 125 & -0.30 $\pm$ 0.12 & 120.86 $\pm$ 0.43 & 2880 $\pm$ 65 \\
K10516096 & 5900 $\pm$ 70 & -0.10 $\pm$ 0.07 & 84.15 $\pm$ 0.36 & 1710 $\pm$ 15\\ 
K10963065 &  6015 $\pm$ 70 & -0.21 $\pm$ 0.07 & 103.61 $\pm$ 0.41 & 2160 $\pm$ 35\\ 
  \hline
\end{tabular}
\tablebib{Observational data: \citet{Mathur2012}.}
\tablefoot{Solar seismologic parameters: $\Delta \nu_{\sun}$ = 134.8 $\pm$
  0.5 $\mu$Hz ; $\nu_{\rm max, \sun}$ = 3034 $\mu$Hz \citep{Thiery2000}.}
\end{table*}

\begin{table*}[ht]
\centering
\caption{Observational sample of binary stars selected for the estimation procedure. }
\label{tab:obs-bin}
\begin{tabular}{lccccccc}
  \hline\hline
Star & $T_{\rm eff}$ (K)& [Fe/H] & $\Delta \nu$ ($\mu$Hz) & $\nu_{\rm max}$
($\mu$Hz) & $M$ ($M_{\sun}$) & $R$ ($R_{\sun}$)&
References\\
\hline
$\alpha$ Cen A & 5847 $\pm$ 27 & 0.24 $\pm$ 0.03 & 105.5 $\pm$ 0.5 & 2410 &
1.105 $\pm$ 0.007 & 1.224 $\pm$ 0.003 & 1, a, A\\
$\alpha$ Cen B & 5316 $\pm$ 28 & 0.25 $\pm$ 0.04 & 161.5 $\pm$ 0.5 & 4100 &
0.935 $\pm$ 0.006 & 0.863 $\pm$ 0.005 & 2, a, A\\
70 Oph A & 5300 $\pm$ 50 & 0.04 $\pm$ 0.05 & 161.7 $\pm$ 0.8 & 4500 & 
0.890 $\pm$ 0.020 & -- & 3, b, B\\
  \hline
\end{tabular}
\tablebib{Asteroseismology data: (1)~\citet{Bouchy2002}; (2)~\citet{Kjeldsen2005};
  (3)~\citet{Carrier2006}. Other observables: (a)~\citet{Porto2008};
  (b)~\citet{Eggenberger2008}. Masses and
  radii values: (A)~\citet{Miglio2005}; (B)~\citet{Eggenberger2008}.}
\tablefoot{Solar seismologic parameters: $\Delta \nu_{\sun}$ = 134.8 $\pm$
  0.5 $\mu$Hz ; $\nu_{\rm max, \sun}$ = 3034 $\mu$Hz \citep{Thiery2000}.}
\end{table*}

\begin{table*}[ht]
\centering
\caption{Mass and radius estimates for the observational sample of
  Tables~\ref{tab:obs-K} and \ref{tab:obs-bin}.}
\label{tab:estim-obs}
\begin{tabular}{lcccc}
  \hline\hline
     &\multicolumn{2}{c}{Standard grid} & \multicolumn{2}{c}{Multi grids}\\
Star & $M$ ($M_{\sun}$) & $R$ ($R_{\sun}$) & $M$ ($M_{\sun}$) & $R$ ($R_{\sun}$)\\
\hline
K3656476 & 1.06$^{+0.03}_{-0.04}$ & 1.30$^{+0.01}_{-0.02}$ &
           1.08$^{+0.02}_{-0.08}$ & 1.30$^{+0.01}_{-0.03}$  \\ 
K6116048 & 0.92 $\pm$ 0.04 & 1.19 $\pm$ 0.02 &
           0.96 $\pm$ 0.04 & 1.20 $\pm$ 0.02 \\
K7976303 & 1.06$^{+0.02}_{-0.04}$ & 1.94 $\pm$ 0.02 & 
           1.08$^{+0.04}_{-0.02}$ & 1.95$^{+0.03}_{-0.02}$\\
K8006161 & 0.95 $\pm$ 0.03 & 0.92 $\pm$ 0.01 & 
           0.96$^{+0.08}_{-0.06}$ & 0.93 $\pm$ 0.02\\
K8379927 & 0.96$^{+0.02}_{-0.03}$ & 1.05 $\pm$ 0.02 &
           0.99 $\pm$ 0.06 & 1.07 $\pm$ 0.02 \\
K10516096 & 1.01 $\pm$ 0.04 & 1.37 $\pm$ 0.02 & 
           1.05$^{+0.05}_{-0.04}$ & 1.39 $\pm$ 0.02 \\
K10963065 & 0.98 $\pm$ 0.04 & 1.18 $\pm$ 0.02 & 
            1.00$^{+0.04}_{-0.05}$ & 1.19 $\pm$ 0.02\\
\hline
$\alpha$ Cen A & 1.09$^{+0.01}_{-0.02}$ & 1.21 $\pm$ 0.01 &
                 1.10$_{-0.02}$\tablefootmark{*} & 1.21 $\pm$ 0.01\\
$\alpha$ Cen B & 0.92$^{+0.02}_{-0.01}$ & 0.86 $\pm$ 0.01 &
                 0.95$^{+0.03}_{-0.04}$ & 0.87 $\pm$ 0.01\\
70 Oph A & 0.86 $\pm$ 0.02 & 0.84 $\pm$ 0.01 &
           0.90$^{+0.02}_{-0.04}$ & 0.85 $\pm$ 0.01 \\
\hline
\end{tabular}
\tablefoot{\tablefoottext{*}{No upper bound because the estimate is at
    the mass grid edge.}}
\end{table*}

\begin{table*}[ht]
\centering
\caption{Mass and radius estimates from literature with different methods.}
\label{tab:estim-ref}
\begin{tabular}{lccccccc}
  \hline\hline
 & \multicolumn{2}{c}{RADIUS} &
  \multicolumn{2}{c}{YB} & \multicolumn{2}{c}{SEEK}\\
Star & $M$ ($M_{\sun}$) & $R$
($R_{\sun}$) & $M$ ($M_{\sun}$) & $R$ ($R_{\sun}$) & $M$ ($M_{\sun}$) & $R$
($R_{\sun}$) & Ref.\\
\hline
K3656476 &  
  1.29 $\pm$ 0.06\tablefootmark{**} & 1.38 $\pm$ 0.02\tablefootmark{**} & 
  1.05 $\pm$ 0.04 & 1.28 $\pm$ 0.02 &
  1.05$^{+0.06}_{-0.03}$ & 1.32 $\pm$ 0.02 & a\\ 
K6116048 & 
  0.86 $\pm$ 0.03 & 1.16 $\pm$ 0.01 & 
  1.03 $\pm$ 0.03\tablefootmark{*} & 1.24$^{+0.01}_{-0.02}$\tablefootmark{*} & 
  0.93$^{+0.07}_{-0.05}$ & 1.19$^{+0.04}_{-0.03}$ & a\\
K7976303 & 
  1.04 $\pm$ 0.03 & 1.93 $\pm$ 0.02 & 
  1.10$^{+0.05}_{-0.08}$ & 2.07 $^{+0.05}_{-0.07}$\tablefootmark{*}  & 
  1.05$^{+0.08}_{-0.04}$ & 1.98$^{+0.03}_{-0.08}$ & a\\
K8006161 & 
  1.07 $\pm$ 0.03\tablefootmark{*} & 0.96 $\pm$ 0.01\tablefootmark{*} & 
  0.96$^{+0.08}_{-0.04}$ & 0.91 $\pm$ 0.01 & 
  1.00 $\pm$ 0.02 & 0.93 $\pm$ 0.02 & a \\
K8379927 & 
  0.84 $\pm$ 0.03\tablefootmark{*} & 1.01 $\pm$ 0.01\tablefootmark{*}  & 
  1.10 $\pm$ 0.06\tablefootmark{*} & 1.13 $\pm$ 0.02\tablefootmark{*} &  
  0.98$^{+0.05}_{-0.08}$ & 1.06 $\pm$ 0.03 & a\\
K10516096 &
  1.00 $\pm$ 0.05 & 1.36 $\pm$ 0.03 &
  1.02 $\pm$ 0.04 & 1.18 $\pm$ 0.03\tablefootmark{**} &
  1.05$^{+0.10}_{-0.05}$ & 1.41 $\pm$ 0.03 & a\\
K10963065 & 
   0.95 $\pm$ 0.04  &  1.17 $\pm$ 0.02 &  
   1.02$^{+0.06}_{-0.07}$ & 1.19 $\pm$ 0.03 &
   1.03$^{+0.07}_{-0.05}$ & 1.21 $\pm$ 0.02 & a\\ 
\hline
$\alpha$ Cen A &  & & & & 
  1.09 $\pm$ 0.09 & 1.23 $\pm$ 0.04 & b \\
$\alpha$ Cen B &  & & & &
  0.92 $\pm$ 0.04 & 0.87 $\pm$ 0.01 & b \\
70 Oph A &  & & & & 
  0.89 $\pm$ 0.06 & 0.86 $\pm$ 0.02 & b\\
\hline
\end{tabular}
\tablefoot{
\tablebib{(a)~\citet{Mathur2012}; (b)~\citet{Quirion2010}.}
\tablefoottext{*}{Difference from standard estimate of
    Table~\ref{tab:estim-obs} greater than $1 \sigma$.}
\tablefoottext{**}{Difference from standard estimate of
    Tab.~\ref{tab:estim-obs} greater than $2 \sigma$.}
}
\end{table*}

\section{Comparison with observations}
\label{sec:observations}

Beyond the tests described in previous sections, any recovery procedure
has to prove its performance against real data.  A severe empirical test is
provided by 
binaries stars for which precise
mass and radius measurements are available.

We focus our attention on three binary stars for which independent estimates
of mass and radius exist. The selected stars are $\alpha$ Cent A + B and 70
Oph A.  These objects have recently been studied by \citet{Quirion2010} with the
SEEK technique, so we can compare our results with both observations and SEEK
estimates. The seismic and non-seismic observable quantities adopted in the
estimation are listed in Table~\ref{tab:obs-bin}.

In Table~\ref{tab:estim-obs} we present the results of the estimation
procedure. This table lists the results obtained using the standard grid alone 
and those obtained also using the grids with
different mixing-length and initial helium abundance values. The
second approach is similar to the one adopted in the SEEK technique.  In all
the six cases the estimates of mass and radius reported in
Table~\ref{tab:estim-obs} are consistent within the
errors.

The corresponding estimates obtained with the SEEK technique are in
Table~\ref{tab:estim-ref}.  It is apparent that the mass and radius obtained
from the recovery grid are consistent 
with the values presented in Table~\ref{tab:obs-bin}.

\section{Conclusions}
\label{sec:conclusions}

In this work we investigated how the stellar model grid-based estimates of
mass and radius of a star are influenced by systematic uncertainties arising
from still uncertain knowledge of both the main input physics
(radiative opacity, nuclear reaction cross sections, etc.) and
macroscopic processes (superadiabatic convection, element diffusion, etc.)
implemented in stellar codes.
 
To do that, we developed a code -- SCEPtER (Stellar CharactEristics Pisa
Estimation gRid), available on-line -- to estimate stellar mass and radius
through a grid-based maximum likelihood technique following
\citet{Gai2011} and \citet{Basu2012}.  In the current version, we relied on
four 
observable quantities, namely the effective temperature, the metallicity
[Fe/H], the large frequency spacing $\Delta \nu$, and the frequency of maximum
oscillation power $\nu_{\rm max}$ of the star. The grid of stellar models
covers the evolutionary phases from ZAMS to the central hydrogen exhaustion in
the mass range [0.8 - 1.1] $M_{\sun}$.

The present work focussed on estimating the statistical errors arising
from the uncertainty in observational quantities and on estimating the
systematic biases due to the uncertainties in initial helium content, in the
mixing-length value, and in some input physics that enter into the stellar
computations. For this last point, we focussed our attention on radiative
opacities, on microscopic diffusion velocities, and on the
$^{14}$N$(p,\gamma)^{15}$O reaction rate. These are the main sources of
uncertainty in the considered evolutionary stages \citep[see][for a detailed
  discussion]{incertezze1, incertezze2}. This is the first time that such an
issue has been addressed.

We found that the statistical error component is almost the same for all the cases we
studied. The standard deviations for mass and radius estimates are about
 4.5\% and 2.2\%, even if the error on the single estimate can reach 
 20\% and 10\%, respectively.

The initial helium content adopted in the stellar computations is assumed to
scale linearly with the metallicity $Z$. To model the uncertainty in its
determination we adopted a reference slope of $\Delta Y/\Delta Z$ = 2 and
studied the effect of choosing different slopes for the synthetic datasets,
namely $\Delta Y/\Delta Z$ = 1 and 3.  The systematic bias due to the
variation in the initial helium content on the explored range is of the order of
$\pm  2.3\%$ on mass and $\pm  1.1\%$ on radius, on average. 
However, for metal-rich stars
(i.e. [Fe/H] $\geq$ 0.0) the effect gets as large as $\pm  4.8\%$ and $\pm  1.9\%$
 for mass and radius estimates, becoming comparable with the
statistical error.

The impact of the mixing-length value variation was studied by computing several
synthetic grids with $\alpha_{\rm ml}$ from 1.50 to 1.98, with our solar
calibrated value (i.e. $\alpha_{\rm ml}$ = 1.74) adopted as a reference for
the recovery standard grid.  We found that the bias induced by the extreme
allowed variation is of the order of $\pm  2.1\%$ on mass and $\pm 1.0\%$ on radius.

The impacts of the uncertainties in input physics that enter into the stellar
computations have been studied here for the first time.  We found that the current
uncertainty in radiative opacities -- i.e. $\pm5\%$ \citep{incertezze1} -- accounts
for a bias of about $\mp  1.0\%$ and $\mp 0.45\%$ in mass and radius determination, whereas
the other explored uncertainty sources in the input physics only have a minor
effect.

The combination of the biases of several sources of uncertainty showed that
they can be directly added with no interaction effects. As an example, the bias due to the
concomitant uncertainty in mixing-length and in initial helium content is about $\pm  4.3\%$
and $\pm 2.0\%$ in mass and radius determination, respectively. These values are very
close to the statistical ones, implying that the estimates can be distorted in
a non-negligible way.

Since several widely used databases of stellar models
\citep[e.g.][]{teramo04,teramo06,padova08,padova09} and some grid-based
technique, such as RADIUS \citep{Stello2009} and SEEK, which both adopt a grid of
models computed with the Aarhus STellar Evolution Code \citep{Dalsgaard2008},
do not implement diffusion, we considered the bias due to this neglect.  We
found that this bias is on average of about 3.7\% and 1.5\%
on mass and radius determination. We also showed that the mass and radius
estimated by relying on a grid of stellar models computed including microscopic
diffusion, but neglecting the temporal evolution of
the surface [Fe/H] in the recovery procedure, are affected by a bias of the order of two-thirds of what is produced
by using stellar tracks without diffusion. In both cases, the bias is greater
at later evolutionary phases and around [Fe/H]~=~$-0.5$.

These results show that the lack of precise knowledge about the physics of the
star might result in biases that are in some cases of the same magnitude of the
uncertainty arising from the observation errors. Since this last source of
uncertainty is expected to shrink in the near future, owing to instrument
improvements, the systematic bias will soon be the main source of uncertainty
in the estimates provided by grid-based techniques. The halving of the
observational errors with respect to those considered in the paper, a goal
already reached for several stars, will decrease the statistical errors  on
mass and radius estimates at the level of their biases.

We compared the results obtained by the SCEPtER technique to those found
with other  
grid-based techniques reported in the literature: RADIUS \citep{Stello2009}, YB
\citep{Gai2011}, and SEEK \citep{Dalsgaard2008}.  Selecting a homogeneous
subset of seven targets from the Kepler catalogue, analysed in
\citet{Mathur2012}, we found that our estimates of mass and radius always
agreed with those of the SEEK technique. Several disagreements were
found in the comparisons with YB estimates, which are obtained with the same
recovering technique used by us but with different stellar models grids.  We
found that the systematic differences among the estimates of the four
techniques are from three to five times greater than the statistical errors of the
estimates. This implies that the differences in the inputs of the stellar
computations are at present the most important source of systematic biases on
the mass and radius estimates obtained by grid-based methods.

Finally, we tested our recovery procedure against three binary stars 
for which mass and radius have been determined empirically.
These systems have been 
studied
by \citet{Quirion2010} with the SEEK technique.  In all three cases our grid-based
estimates of mass and radius agree with both the observations
and the estimates obtained with the SEEK technique.

\begin{acknowledgements}
We are grateful to our anonymous referee for many stimulating suggestions that
help

 clarify and improve the paper.
We thank Steven N. Shore for carefully reading the paper and for
useful comments. 
This work has been supported by PRIN-INAF 2011 
 ({\em Tracing the formation and evolution of the Galactic Halo with VST}, PI
M. Marconi) and PRIN-MIUR 2010-2011 ({\em Chemical and dynamical evolution 
of the Milky Way and Local Group galaxies}, PI F. Matteucci).

\end{acknowledgements}

\bibliographystyle{aa}
\bibliography{biblio}

\appendix

\section{Kernel density and LOWESS}
\label{sec:smooth}

The kernel density is a non-parametric estimate of the probability density
function from a discrete set of data. It can be viewed as a generalization of
the histogram, with better theoretical properties \citep{simar}. For a set of
$n$ observations $x_1$, $x_2$, $\ldots$, $x_n,$ a kernel density with bandwidth
$h$ has the form:
\begin{equation}    
\hat f(x,h) = \frac{1}{n h} \sum_{i=1}^n K\left(\frac{x-x_i}{h}\right)
\end{equation}    
where the kernel function $K$ is chosen to be a probability density function.
Several choices of kernel are available. In this work. we make use of a
Gaussian kernel:
\begin{equation}    
K(y) = \frac{1}{\sqrt{2 \pi}} \exp\left( - \frac{y^2}{2}\right).
\end{equation}
The kernel selection usually has a minor influence on the kernel estimate with
respect to the bandwidth $h$. This parameter is selected balancing two effects
since an increment of $h$ increases the bias of $\hat f$, while it reduces its
variance. An often adopted choice for a Gaussian kernel is given by the rule of
thumb \citep{silverman1986density}:
\begin{equation} 
\hat h = 0.9 \; {\rm min}(\hat \sigma, R/1.34) \; n^{-1/5}
\end{equation}
where $\hat \sigma$ is the sample standard deviation and $R$  the sample
interquartile range.

Other choices for the bandwidth, based on the asymptotic expansion of the mean
integrated squared error, are reported in the literature.  The different
choices have an impact on kernel estimator for multi-modal
distributions, which is not the case of the present work.  We refer
interested readers to \citet{Feigelson2012, simar, venables2002modern,
  Sheathe1991}.

A frequently used bivariate smoother is the local regression technique LOWESS,
which combines a linear least squares regression with a robust nonlinear
regression.  It provides a generally smooth curve, whose value at a
particular location along the $x$ axis is only determined by the points in its
neighbourhood.  The first step is to fit a polynomial regression in a
neighbourhood of $x$. A fraction $f$ of the $n$ sample points near $x$ is
selected. We define $m = \lceil f n \rceil$ the number of points used in
the fit. Then the technique obtains the estimates $\hat \beta$ by minimizing
\begin{equation}
\hat y = m^{-1} \sum_{i=1}^m W_{i}(x) \left(y_i - \sum_{j=0}^p \beta_j
x_j\right)^2 \; ,
 \label{eq:lowess}
\end{equation}
where $W_{i}(x)$ are the weights, usually obtained by the tricubic function:
\begin{equation}
W_{i}(x) = \left(1-\left|\frac{x-x_i}{d}\right|^3\right)^3 \; ,
\end{equation}
where $d$ is the maximum distance between $x$ and the other predictor values
$x_i$ in the span.  For LOWESS estimates the value $p = 1$ is usually adopted,
implying a local linear regression.  The model residuals $\hat \epsilon_i$ and
the scale parameter $\hat m = {\rm median} (\hat \epsilon_i)$ are computed.
The median absolute deviation $\hat \sigma$ of the residuals is evaluated:
$\hat \sigma = {\rm median} |\hat \epsilon_i - \hat m|$.  Then the algorithm
computes the robustness weights $\delta_i = R(\hat \epsilon_i/6 \hat \sigma)$,
with $R(u) = 15/16 (1-u)^2$ for $|u| \leq 1$ and $R(u) = 0$ otherwise.  The
local regression of Eq.~(\ref{eq:lowess}) is computed again, but with weights
given by $\delta_i K_i(x)$. This procedure is iterated a variable number of
times between one and five; this makes the local estimate robust even in
presence of outliers.  Further details on the technique and on the numerical
methods used to speed up the computations are available in
\citet{Cleveland1981}.

The computations outlined in this section were performed using the functions
{\it density} and {\it lowess}, available in the R 2.15.2 \citep{R}.

\end{document}